\documentclass[12pt,preprint]{aastex}

\begin{document}

\shortauthors{Taniguchi et al.}
\shorttitle{HST/ACS morphology of Ly$\alpha$ emitters at $z = 5.7$}


\title{HST/ACS MORPHOLOGY OF LYMAN $\alpha$ EMITTERS AT REDSHIFT 5.7 IN THE COSMOS FIELD\altaffilmark{1}}

\author{
Y.~Taniguchi       \altaffilmark{2},
T.~Murayama        \altaffilmark{3},
N.~Z.~Scoville     \altaffilmark{4},
S.~S.~Sasaki       \altaffilmark{3,5},
T.~Nagao           \altaffilmark{2,5},
Y.~Shioya          \altaffilmark{2},
T.~Saito           \altaffilmark{2},
Y.~Ideue           \altaffilmark{5},
A.~Nakajima        \altaffilmark{5},
K.~Matsuoka        \altaffilmark{5},
D.~B.~Sanders      \altaffilmark{6},
B.~Mobasher        \altaffilmark{7},
H.~Aussel          \altaffilmark{8},
P.~Capak           \altaffilmark{4,9},
M.~Salvato         \altaffilmark{4},
A.~Koekemoer       \altaffilmark{10},
C.~Carilli         \altaffilmark{11},
A.~Cimatti         \altaffilmark{12},
R.~S.~Ellis        \altaffilmark{4},
B.~Garilli         \altaffilmark{13},
M.~Giavalisco      \altaffilmark{14},
O.~Ilbert          \altaffilmark{6},
C.~D.~Impey        \altaffilmark{15},
M.~G.~Kitzbichler  \altaffilmark{16},
O.~Le~Fevre        \altaffilmark{17},
H.~J.~McCracken    \altaffilmark{18},
C.~Scarlata        \altaffilmark{4},
E.~Schinnerer      \altaffilmark{19},
V.~Smolcic         \altaffilmark{4},
S.~Tribiano        \altaffilmark{20}, and
J.~R.~Trump        \altaffilmark{15}
		}

\altaffiltext{1}{Based on observations with NASA/ESA {\it Hubble Space Telescope}, obtained at
the Space Telescope Science Institute, which is operated by AURA, Inc., under NASA contract
NAS 5-26555; and also based on data collected at 
	Subaru Telescope, which is operated by 
	the National Astronomical Observatory of Japan.}
\altaffiltext{2}{Research Center for Space and Cosmic Evolution, Ehime University, 
        Bunkyo-cho 2-5, Matsuyama 790-8577, Japan}
\altaffiltext{3}{Astronomical Institute, Graduate School of Science,
        Tohoku University, Aramaki, Aoba, Sendai 980-8578, Japan}
\altaffiltext{4}{Department of Astronomy, MS 105-24, California Institute of
                Technology, Pasadena, CA 91125}
\altaffiltext{5}{Graduate School of Science and Engineering, Ehime University, 
        Bunkyo-cho, Matsuyama 790-8577, Japan}
\altaffiltext{6}{Institute for Astronomy,  University of Hawaii,
                 2680 Woodlawn Drive, HI 96822}
\altaffiltext{7}{Department of Physics and Astronomy, University of California,
                 Riverside, CA 92521}
\altaffiltext{8}{CEA/DSM-CNRS, Universit\'e Paris Diderot, DAPNIA/SAp,
                 Orme des Merisiers, 91191 Gif-sur-Yvette, France}
\altaffiltext{9}{Spitzer Science Center, California Institute of Technology, Pasadena, CA 91125}
\altaffiltext{10}{Space Telescope Science Institute, 3700 San Martin Drive,
                 Baltimore, MD 21218}
\altaffiltext{11}{National Radio Astronomy Observatory,
                  P.O. Box 0, Socorro, NM 87801-0387}
\altaffiltext{12}{Dipartamento di Astronmia, Universita  di Bologna, Italy}
\altaffiltext{13}{INAF, Istituto di Astrofisica Spaziale e Fisica Cosmica,
                  Sezione di Milano, via Bassini 15, 20133 Milano, Itary}
\altaffiltext{14}{Department of Astronomy, University of Massachusetts, Amherst, MA 01003}
\altaffiltext{15}{Steward Observatory, University of Arizona, Tucson, AZ 85721}
\altaffiltext{16}{Max-Planck-Institut f\"ur Astrophysik,
                  D-85748 Garching bei M\"unchen, Germany}
\altaffiltext{17}{Laboratoire d'Astrophysique de Marseille,
                  BP 8, Traverse du Siphon, 13376 Marseille Cedex 12, France}
\altaffiltext{18}{Institut d'Astrophysique de Paris, 98 bis Boulevard Arago, F-75014 Paris, France}
\altaffiltext{19}{Max Planck Institut f\"ur Astronomie,
                  K\"onigstuhl 17, Heidelberg, D-69117, Germany}
\altaffiltext{20}{City University of New York, Borough of Manhatan Community College
                  199 Chambers St., New York, NY 10007}

\begin{abstract}

We present detailed morphological properties of Ly$\alpha$ emitters (LAEs) at $z\approx 5.7$
in the COSMOS field, based on {\it Hubble Space Telescope} Advanced Camera
for Surveys (ACS) data. The ACS imaging in the F814W filter covered 85 LAEs of the 119 LAEs identified
in the full two square degree field, and 47 LAEs of them are detected in the ACS images.
Nearly half of them are spatially extended with a size larger than 0.15 arcsec ($\sim$0.88 kpc at $z=5.7$)
up to 0.4 arcsec ($\sim$2.5 kpc at $z=5.7$).
The others are nearly unresolved compact objects.
Two LAEs show double-component structures,
indicating interaction or merging of building components to form more massive galaxies.
By stacking the ACS images of all the detected sources,
we obtain a Sersic parameter of $n \sim 0.7$
with a half-light radius of 0.13 arcsec (0.76 kpc),
suggesting that the majority of ACS detected LAEs have not spheroidal-like but
disk-like or irregular light profiles.
Comparing  ACS F814W magnitudes ($I_{814}$) with Subaru/Suprime-Cam magnitudes in the $NB816$, $i'$, and $z'$ bands,
we find that the ACS imaging in the F814W band mainly probes UV continuum rather than
Ly$\alpha$ line emission.
UV continuum sizes tend to be larger for LAEs with larger Ly$\alpha$ emission regions
as traced by the $NB816$ imaging.
The non-detection of 38 LAEs in the ACS images is likely due to the fact
that their surface brightness is even too low both in the UV continuum and Ly$\alpha$ emission.
Estimating $I_{814}$ for the LAEs with ACS non-detection from the $z'$ and $NB816$ magnitudes,
we find that 16 of these are probably LAEs with a size larger than 0.15 arcsec in UV continuum.
All these results suggest that our LAE sample contains systematically larger LAEs
in UV continuum size than those previously studied at $z\sim6$.

\end{abstract}

\keywords{cosmology: observations ---
   cosmology: early universe ---
   galaxies: formation ---
   galaxies: evolution ---
   galaxies: morphology}

\section{INTRODUCTION}

During the last decade, a large number of young, star-forming galaxies beyond 
redshift of $z=5$ have been found based on deep imaging 
observations of both the {\it Hubble Space Telescope} (HST) and
8-10m class telescopes such as the 8.2m Subaru Telescope
(see for a review, Bouwens \& Illingworth 2006; Taniguchi 2008;
see also for recent progress, Bouwens et al. 2008, 2009; Bradley et al. 2008).
In particular, narrow-band imaging surveys have been providing us
well-defined samples of strong Ly$\alpha$ emitters (LAEs) at $z\approx 5.7$
(e.g., Rhodes \& Malhotra 2001; Ajiki et al. 2003; Hu et al. 2004;
Shimasaku et al. 2006; Murayama et al. 2007 and references therein). 
These surveys are used to investigate the star formation activity in such young galaxies,
providing typical star formation rates from several to a few tens $M_{\odot}$ yr$^{-1}$.
Clustering properties are also one of the important issues studied as structure
formation in the early universe provides an important observational constraint on
hierarchical structure formation scenarios (e.g., Springel et al. 2005).
Although a possible over-density region has been identified in the Subaru XMM-Newton Deep Survey
Field (Ouchi et al. 2005), there is no other significant evidence for clustering of 
young galaxies at $z\approx 5.7$ in other deep wide-field surveys
(e.g., Shimasaku et al. 2006; Murayama et al. 2007).  

Another interesting question is addressed to morphological properties of LAEs because 
these properties give us insights on  how LAEs were assembled
and how their intense star formation events were triggered.
However, no systematic investigation of the detailed morphology of high-$z$ LAEs
has yet been undertaken although some case studies have been reported
(e.g., Rhoads et al. 2005; Venemans et al. 2005; Pizkal et al. 2007;
Overzier et al. 2008 and references therein).
Rhoads et al. (2005) found in the Hubble Ultra  Deep Field (UDF)
a LAE at $z=5.4$ (UDF 5225) with a spatial
extent of 0.25 arcsec $\times$ 1.0 arcsec (1.6 kpc $\times$ 6.3 kpc).
This LAE shows a core together with three knots that appear to emanate from
the core.  On the other hand, Bunker et al. (2003) found a very compact LAE
at $z=5.78$ in the {\it Chandra} Deep Field South among their $i$-dropout sample.
Its half light radius is only 0.08 arcsec (490 pc), barely resolved by the ACS PSF (0.06 arcsec).
Stanway et al. (2004a) found three LAEs in the UDF during the course of their
GLARE project (= Gemini Lyman Alpha at Reionization Era) and obtained also small 
sizes ranging from 0.09 arcsec ($\sim$ 500 pc) to 0.14 arcsec ($\sim$ 900 pc).
Relative small sizes of six LAEs with respect to
Lyman break galaxies (LBGs) at similar redshift ($z \sim 6$) are also found
by Dow-Hygelund et al. (2007).  
Since the size and morphological properties provide us with useful insight on the
understanding of physical processes of star formation,
we need systematic studies of the detailed morphological properties for 
a large sample of such high-$z$ galaxies.

Recently, Murayama et al. (2007; hereafter M07) identified a total of 119
LAEs at $z\approx 5.7$ in the Cosmic Evolution Survey (COSMOS) field (Scoville et al. 2007a),
providing one of the largest samples of LAEs in a large contiguous field.
Since F814W imaging taken with the Advanced Camera for Surveys (ACS)
on-board the {\it HST} is available for the COSMOS field
(Scoville et al. 2007b; Koekemoer et al. 2007), 
the sizes and morphologies of the LAEs in the COSMOS field 
can be investigated in detail.
In particular, since the redshifted Ly$\alpha$ emission of our LAE sample
is probed in the F814W imaging, it is possible to study the Ly$\alpha$ morphology
as well as the rest-frame ultraviolet continuum shape.
In this paper, we present our detailed analysis of ACS images of the LAE sample of M07.

We use a standard cosmology with 
$\Omega_{\rm matter} = 0.3$, $\Omega_\Lambda = 0.7$,
and $H_0 = 70$ km s$^{-1}$ Mpc$^{-1}$.
Throughout this paper, we use magnitudes in the AB system.

\section{OBSERVATIONAL DATA AND ACS COUNTERPARTS OF LAEs}

In M07, 119 LAE candidates at $ 5.65 < z < 5.75$
were carefully selected from optical imaging with both the narrow-band filter, NB816
($\lambda_c = 8150$ \AA~ with a width $\Delta\lambda =120$ \AA~ 
see Ajiki et al. 2003 for details) and broad-band filters from $B$ to $z^{\prime}$ 
taken for a 1.95 deg$^{2}$ area of the COSMOS field using
Suprime-Cam (Miyazaki et al. 2002) on the Subaru Telescope
(Kaifu et al. 2000; Iye et al. 2004).
Details of the Subaru observations and data processing are
described by Taniguchi et al. (2007) and Capak et al. (2007).
Note that follow-up spectroscopy has been performed
for 24 LAEs in the sample and that all of them showed Ly$\alpha$ emission
at $z\approx 5.7$ (Capak et al. 2009) verifying the effectiveness of the adopted selection method.

The ACS data in the F814W filter were taken for 
an area of 1.64 deg$^{2}$ of the COSMOS field
and were processed to 0.05$^{\prime\prime}$ pixel$^{-1}$ images with an
averaged point spread function (PSF) width of 0.097$^{\prime\prime}$
(Scoville et al. 2007b; Koekemoer et al. 2007).
In our analysis, we use the official COSMOS ACS image, Version 1.3.

Since the observed area of the ACS imaging is slightly smaller than
that of the Subaru one, we find that
ACS data are available for 85 LAEs in the LAE sample defined by M07.
Spatial distribution of these LAEs is shown with red symbols in Fig. 1.
Among them, 20 LAEs are already spectroscopically identified as star-forming galaxies 
at $z \sim 5.7$ (Capak et al. 2009).


Our data analysis procedures for ACS data are as follow.
First, for each LAE, we created a small ACS cut-out
 (16$^{\prime\prime}$$\times$16$^{\prime\prime}$) centered on the 
the LAE position (as derived in the  NB816 image; see M07). 
Then, we used SExtractor (Bertin \& Arnouts 1996)
to detect a possible ACS counterpart of the LAE.
We adopted the following SExtractor parameters: a detection threshold of 1.6$\sigma$ and 
a minimum detection area of 9 pixels.
These values were carefully selected (1) to allow for the detection both diffuse elongated sources as well as 
point-like sources, (2) to minimize false detections, and (3)
to minimize detection failures for sources  that are apparently shown in the ACS image by eyes.   
A gaussian-profile filter (5 pixels $\times$ 5 pixels) with a FWHM
of 2 pixels (0.1$^{\prime\prime}$) well matched to the PSF size was applied
to the detection images for smoothing.
The SExtractor parameters for the source detection in $I_{814}$ are listed in  Table 1.

We found 58 sources detected near the LAE positions ($r \le 1^{\prime\prime}$).
By eye inspection  we rejected 3 sources because they are largely offset 
($\ge0.68^{\prime\prime}$) from the LAE center
and have no distinct counterpart in the Subaru broad-band images of any bands.
Nest, in order to remove low-$z$ foreground neighbors, we examined 
the Subaru  $B$, $V$, and $g'$ images.
Since the observed wavelength of the Lyman limit at $z\approx 5.7$ shifts to 6110 \AA{}, 
a true LAE must be undetectable in these bands.
In this analysis, we adopted the search radius for ACS counterparts is 1$^{\prime\prime}$
and then removed 6 sources.

Finally, among the 85 LAEs whose ACS data are available,
we found 49 ACS sources for 47 LAEs (17 objects were spectroscopically confirmed;
Capak et al. 2009).
Two LAEs (No.~60 and No.~110) have double-component ACS sources.
Offsets of the ACS positions from the NB816 positions are typically less than 0.13$^{\prime\prime}$
in the ACS images,
smaller than the pixel scale (0.15$^{\prime\prime}$ pixel$^{-1}$) of the NB816 images.
The remaining 38 LAEs (3 objects were spectroscopically confirmed; Capak et al. 2009) 
are not detected in the ACS images. 
The numbers of the total sample, both the detected and not-detected LAEs are summarized in Table 2.

In Fig. 2, we show thumbnails of the 85 LAEs in the ACS F814W images. 
We also show thumbnails in the ACS F814W images 
together with their Subaru $i^{\prime}$, $NB816$, and  $z^{\prime}$ images in Fig. 3.
Smoothed F814W images with a gaussian kernel with a FWHM of 2 pixels are also presented.
The detected sources identified as LAE counterparts are indicated by red ellipses on the smoothed ACS, 
$NB816$, and  $z^{\prime}$images.
Green and blue ellipses are ACS sources excluded from the sample by eye inspection and by rejection of  
foreground neighbors, respectively.

As shown in Fig. 1, the ACS detected  (red filled circles) and  not-detected LAEs (red crosses)
appear to be  almost randomly distributed in the COSMOS field and thus 
their distributions may not be affected by large-scale inhomogeneity
of the ACS data quality (e.g., edges of the field).
The total magnitude ($I_{814}$) and half-light radius ($R_{\rm HL}$) were measured for each 
detected source by SExtractor
on the original ACS image (i.e., not on the smoothed image).
We adopted SExtractor's MAG\_AUTO as total magnitude.
For the 38 LAEs undetected in the F814W image, we estimated 3$\sigma$
upper-limits for the magnitudes within a 1$^{\prime\prime}$ diameter aperture.
These photometric properties of the ACS data are listed in Table 3. 
Note that the 3$\sigma$ limiting magnitude of the  F814W images is 27.3 mag in
a 1$^{\prime\prime}$ diameter aperture.
All magnitudes are corrected for the Galactic extinction of $A_{F814W}$=0.035 (Capak et al. 2007).
In Table 4, we list the photometric properties of the LAE candidates from M07.
The 3$\sigma$ limiting magnitudes  within a 2$^{\prime\prime}$ diameter aperture in 
the NB816, $i'$, and  $z'$ images are 25.7, 26.1, and 25.3, respectively.





\section{MORPHOLOGICAL PROPERTIES}

\subsection{Half Light Radius}

The ACS counterparts of the LAEs look differently from object to object as shown in Fig. 2.
Some LAEs show compact, round, and nearly unresolved  shapes, while
others have an elongated, filamentary, or patchy morphology.

First, we analyze the sizes of our LAE sample in the ACS images.
We show the  distribution of half-light radius  ($R_{\rm HL}$) 
for the 47 ACS-detected LAEs in Fig. 4.
In this histogram, the size of the larger component is adopted for the two LAEs with 
double-component ACS sources.
Note that the measured half-light radii of stars in the ACS images are typically 0.11$^{\prime\prime}$
as indicated in Fig. 4.
The number of LAEs in each $R_{\rm HL}$ bin decreases with increasing  half-light radius beyond 0.15$^{\prime\prime}$
while the overall distribution appears to be almost flat at the radius of $R_{\rm HL} < 0.15^{\prime\prime}$.
Therefore, we may conclude that  LAEs with $R_{\rm HL} < 0.15^{\prime\prime}$ are almost unresolved compact objects.
Three LAEs (Nos. 40, 68, and 78) have a larger size as $R_{\rm HL} \ge  0.3^{\prime\prime}$
compared with the PSF size.
However, the sizes of all the LAEs detected in our ACS imaging are smaller than $0.4^{\prime\prime}$ and 
no widely extended LAE (like the LAE found by Rhoads et al. 2005) has been found.

 
In Fig. 5, we show the distribution of  $I_{814}$.
This  distribution is affected by detection incompleteness toward
fainter magnitudes because the detection is limited by surface brightness.
For the LAEs with no detection in ACS, we show 
3$\sigma$ upper limits within a 1$^{\prime\prime}$ diameter aperture
in Fig. 6. If we adopt 1$\sigma$ and 2$\sigma$ upper limits,
the limiting magnitudes are fainter by 1.2 and 0.45 magnitude, respectively.




In order to examine the effect of limiting surface brightness in our ACS imaging,
we show the relation between $R_{\rm HL}$ and $I_{814}$ magnitude in Fig. 7.
In this diagram, we find no faint object with a large  $R_{\rm HL}$.
This suggests that we are unable to detect LAEs with $R_{\rm HL} >$ 0.5 arcsec
if they are fainter than 26 mag in $I_{814}$.

In Fig. 8, the frequency distributions of sizes in NB816 imaging, 
$FWHM$(NB816), are shown for the detected and not-detected LAEs.
Not a small part of the sample LAEs have larger sizes than the PSF size (0.98 arcsec),
especially for the not-detected LAEs.
Their appearance in the NB816 images shows widely extended morphology
(e.g., No.~74 in the ACS detected subsample, and No.~1, No.~27, and No.~29
in the ACS not-detected subsample) and looks like a so-called Ly$\alpha$ blob
(e.g., Fynbo et al. 1999; Steidel et al. 2000; Matsuda et al. 2004; Saito et al. 2006).
In Fig. 9, we also show the diagram between $FWHM$(NB816) and $R_{\rm HL}$.
This tendency is more conspicuous for the not-detected LAEs. 
Even if a LAE has a larger size in NB816, i.e., it has an extended Ly$\alpha$ source,
we would be able to detect it in our ACS F814W imaging
if it has a compact UV continuum component less than 0.5 arcsec.
Therefore, it is suggested that the not-detected LAEs have an extended UV continuum
component larger than 0.5 arcsec.
However, the $z^{\prime}$ magnitudes of the not-detected LAEs
are typically fainter relative to the ones detected in ACS (see Table 3).
This means that their UV magnitudes  are in some cases too faint to be detected in our ACS imaging
even though they are not very much extended (i.e., $< 0.5$ arcsec);
see section 3.4 for more details.




\subsection{Monte Carlo Simulations}

To estimate practical errors in our measurements of $R_{\rm HL}$, 
we performed Monte Carlo simulations. 
We prepared 1000 artificial sources with the exponential light profile
for each set of given parameters of the total magnitude, the half light
radius, and the ellipticity. We put these sources on the observed ACS
image after they were convolved with the PSF image. We also added photon noises.
Then, we measured their photometric properties with the same SExtractor 
parameters for the source detection.
Detection completeness of the 50\% limit is indicated in Fig. 7.

Based on these simulations, we estimated probability distributions
of each parameter for each LAE.
The median values with the  68\% confidence intervals of estimated distributions
for total magnitude and half light radius are listed in Table 3.
In the upper panel of Fig. 10, we show the relation between estimated magnitudes
and measured magnitudes.
The estimated magnitudes tend to be brighter than the measured magnitudes.
The estimated errors are typically $\pm 0.3$ --  $\pm 0.5$ and larger
than the measured error (typically $\pm 0.1$--$\pm 0.2$).
Although the estimated  half light radius suffers from large uncertainty,
the LAEs measured with $R_{\rm HL} \ge 0.3^{\prime\prime}$ still have
a large estimated value in $R_{\rm HL}$. 
Moreover, the simulation indicates several LAEs are intrinsically large 
even their measured sizes are nearly the PSF size.
While most of LAEs are unresolved objects,
we may conclude that some of LAEs at $z\sim 5.7$ have larger
intrinsic $R_{\rm HL}$ than the PSF size.


\subsection{What Do We See in ACS F814W Images; Ly$\alpha$ Emission or Ultraviolet Continuum ?}

As we see in Section 3.1, the 47 LAEs detected in our ACS F814W imaging are not largely extended
(i.e., $R_{\rm HL} < 0.4$ arcsec). What do we see exactly in the ACS images?
Here we examine whether the detected light is from Ly$\alpha$ emission or Ultraviolet (UV) continuum.

First, we present the transmission curves for the filters used in our analysis; F814W for ACS/HST
and $i'$, $NB816$, and $z'$ for Suprime-Cam/Subaru in Fig. 11.
The CCD sensitivity is taken into account for each filter transmission curve .
The bandpass of the NB816 filter is  almost centered in the wavelength range cover the F814W filter.
Therefore, if Ly$\alpha$ emission is strong enough to be detected in our F814W imaging,
we would see Ly$\alpha$ morphologies of LAEs. However, the UV continuum 
at wavelengths longer than 1216 \AA{} can also be probed by the F814W imaging
whose transmission curve is similar to the sum of Suprime-Cam $i'$ + $z'$ filter
transmission. In order to investigate what our F814W imaging probes, we examine the correlation
between $I_{814}$ and $NB816$ in Fig. 12.
Since the correlation  appears to be poor (its correlation coefficient is $r=0.40$),
the F814W imaging does not  primarily probe Ly$\alpha$ emission.
Next, in Fig. 13,  we show the correlations between $I_{814}$ and $i'$ ($r=0.32$), and $I_{814}$ and $z'$  ($r=0.45$).
These comparisons show that $I_{814}$ is more correlated with $z'$.
This suggests that the F814W imaging probes UV continuum from massive stars in each LAE,
because  $z'$ is only sensitive to UV continuum as evident from the filter curves (Fig. 11).




Although the correlation between $I_{814}$ and $z^{\prime}$ is better than for $NB816$ and $i'$,
there is a systematic offset between $I_{814}$ and $z'$ values. 
$I_{814}$ is typically 0.94 mag (average) fainter than $z'$.
This offset can be explained by the difference of 
their filter transmission curves  between $I_{814}$ and $z'$ (see Fig. 11).
Since the Ly$\alpha$ wavelength,  1216 \AA{}, is observed  nearly at the band center of the F814W filter,
the UV continuum at wavelengths longer than  1216 \AA{} is covered by 
only half of the F814W bandpass (see Fig. 11).
As a result, flux densities of the UV continuum are underestimated with our F814W imaging 
by roughly 0.75 (=$2.5\log 2$) mag.
For a more precise treatment, we can estimate the  correction factor that converts $I_{814}$ to  $z'$ for each LAE
by assuming a simple model spectrum similar to the one shown in Fig. 11, in which $f_{\nu}$ is assumed to be constant.
Given both the UV continuum flux estimated by our $z^{\prime}$ magnitude
and the rest-frame $EW$(Ly$\alpha$), we estimate the correction factor for each LAE and then
estimate $z^{\prime}$ magnitude based on $I_{814}$ value, $z^{\prime}$($I_{814}$).
For the LAEs with only lower limits of $EW$(Ly$\alpha$),
we use the lower limit values for this estimate.
In Fig. 14, we compare  $z^{\prime}$($I_{814}$) with $z^{\prime}$, and
find that there is a tighter correlation ($r=0.58$) between these two magnitudes.
Therefore, we conclude that we see the UV continuum of LAEs and miss almost all
the Ly$\alpha$ flux in our ACS imaging.


\subsection{Spatial sizes of the LAEs without ACS counterparts}

It is reminded that 38 LAEs are not detected in our ACS imaging. This means that
the surface brightness of these LAEs is too low to be detected not only in Ly$\alpha$ 
but also in their UV continuum. Even for LAEs with bright UV continuum,
they could not be detected in the ACS image if they are
spatially extended and thus their surface brightness falls below the detection limit.
Therefore, some of the LAEs undetected in the ACS images may have large $R_{\rm HL}$.

Here, assuming that the LAEs without ACS counterparts also have the same SED
as that adopted in the previous subsection, we estimate the expected $I_{814}$ magnitude,
$I_{814}$($z^{\prime}$), based on the Subaru $z^{\prime}$ and the $NB816$ magnitudes.
We find that 16 of the 38 not-detected LAEs in ACS would be bright as much as
$I_{814}$($z^{\prime}$) $\le$ 26.5.
For LAEs with $I_{814}$=26.5 and  $R_{\rm HL}$=0.15 arcsec,
the detection completeness is 50\% (see Fig. 7).
Therefore, these 16 LAEs with $I_{814}$($z^{\prime}$) $\le$ 26.5
are probably also slightly extended LAEs ($R_{\rm HL}$ $\ge$ 0.15 arcsec).
However, there is no LAE without ACS counterparts brighter than 25 mag in  $I_{814}$($z^{\prime}$).
This suggests that very extended LAEs with $R_{\rm HL}$ $\ge$ 0.4 arcsec are not present
in our LAE sample.

As shown in the previous subsection, $z^{\prime}$ is well correlated with  $I_{814}$ and
these magnitudes represent UV continuum brightness for the ACS-detected LAEs. 
For the not-detected LAEs, it is not uncertain if this relation would be still valid.
Their $z^{\prime}$ are systematically fainter than those
of the ACS detected LAEs while the magnitude ranges of $NB816$ for both the subsamples are nearly the same,
implying that contribution of  Ly$\alpha$ flux would be considerable for that of the not-detected LAEs.
Since the sizes in $NB816$ of the not-detected LAEs are systematically larger than those of 
the ACS-detected LAEs (Fig. 8),  
it is suggested that a faint compact UV continuum source with a widely extended bright
Ly$\alpha$ nebula explains for the undetection in  $I_{814}$ for a part of the not-detected LAEs.
However, at least for the 16 not-detected LAEs with  $I_{814}$($z^{\prime}$) $\le$ 26.5,
their rest $EW$(Ly$\alpha$) is comparable with that of ACS-detected LAEs, so that
they would be detected in ACS if they had a compact ($R_{\rm HL}$ $<$ 0.15 arcsec) 
UV continuum source. Therefore, they probably have an extended UV continuum source.

\section{DISCUSSION}

\subsection{ACS Size and Star Formation Properties}

In the previous section, we have demonstrated that our ACS imaging with F814W
does not probe Ly$\alpha$ emission but UV continuum from massive stars in the LAEs.
Since all LAEs detected in the ACS imaging are spatially small  (i.e., $<$ 0.4 arcsec or 
$<$2.5 kpc),
their star-forming regions are considered to be physically compact.
The sizes of high-$z$ galaxies provide important information on
the growth of the luminous parts of galaxies embedded in dark matter haloes
(e.g., Dalcanton, Spergel, \& Summers 1997; Mo, Mao, \& White 1998; 
Ferguson et al. 2004; Bouwens et al. 2006). 

Here we investigate how the size of the UV continuum is related to the global star 
formation properties in the LAEs. 
In Fig. 15, we show the diagram between $R_{\rm HL}$ and $L$(Ly$\alpha$). 
In Fig. 16, we present the diagram between $R_{\rm HL}$ and the rest-frame
$EW$(Ly$\alpha$). We find no correlation in these two diagrams.
These results suggest that the size of UV continuum regions is not directly
related to the star-forming activity traced by Ly$\alpha$ emission in our LAEs. 
However, one would naturally have expected positive correlations 
in these diagrams.
Therefore, the little correlation implies the following
three possibilities.
(1) Most star-forming regions have lower surface brightness than our 
detection limit.
(2) Most star-forming regions are hidden by dusty clouds.
Or, 
(3) the UV continuum probes star forming regions with a large age spread
while Ly$\alpha$ only probes youngest star forming regions that might not spatially overlap with 
older star forming regions.
Radio and millimeter stacking analysis for our LAE sample rules out  
large dust content (Carilli et al. 2007).
Since a number of observations also suggest that LAEs at high redshift
tend to have little dust content (Lai et al. 2007; Finkelstein et al. 2008, 2009a, \& 2009b;
see, however, Chary et al. 2005), the second possibility
appears to be  unlikely.
Currently, however, we have no firm answer on this issue.
In future, we need deeper F814 imaging of our LAEs. Also, sensitive
rest-frame, mid- and far-infrared imaging will be necessary to solve this problem.



\subsection{Structural Properties of the LAEs Detected in our ACS Imaging}

As we have shown, our ACS imaging with F814W probes  UV continuum
from massive stars in the LAEs and
the majority of our LAEs show extended morphology in $I_{814}$.
It is of great interest to analyze not only their sizes  but also
their surface brightness profiles.
Unfortunately, our ACS images are not deep enough to perform such an analysis for each LAE individually.
Therefore, we alternatively adopt the image stacking method
to obtain mean properties for our LAEs at $z\approx 5.7$.

First, we rotated  the major axis to align in the $x$-axis for each LAE image.
In this procedure, we used the position angles measured by SExtractor.
We also constructed second rotated images in 180 degree from the first rotated (major-axis aligned) images.
Next, we generated a composite image by co-adding counts of  both the  first and second rotated images of LAEs.
In this analysis, we did not use the double-component LAEs, No. 60 and No. 110.
We also excluded No. 73 and No. 81 because foreground galaxies
are very close to these LAE sources.
Therefore, we used the remaining 43 LAEs for the stacking analysis.

Finally, we measured structural parameters on the resultant image by modeling
1-dimensional surface brightness profile along the major axis.
The PSF image required for convolving the model light distribution was
derived by combining stellar images near the LAEs.
Assuming a Sersic function, we obtained the PSF-deconvolved effective radius of
$R_{\rm HL}=0.13^{+0.03}_{-0.01}$ arcsec  (0.76 kpc) and the 
Sersic index ($n$) of $n=0.7^{+0.3}_{-0.3}$.
In Fig. 17, we show the composite images, light profiles of the composite image along the major axis, 
and the model light profile.
Note that the observed effective radius of stars are typically 0.11 arcsec,
and thus the composite LAE image is  only slightly resolved.
The derived parameters may contain larger uncertainties than the nominal errors estimated from the residuals.
However, the derived small Sersic index suggests that our LAEs may have on average irregular or disk-like morphologies
rather than spheroidal structures.

Ravindranath et al. (2006) found 40\% of Lyman break galaxies (LBGs) at $z > 2.5$ have exponential profiles,
30\% of LBGs have $r^{1/4}$-like profiles, and 30\% of LBGs have multiple cores.
In contrast, our analysis indicates that spheroidal structures are rare in our LAE sample at $z\approx5.7$.
Note that clumpy- or chain-like structures are often seen in galaxies at $z$=1--5 in the UDF sample
(e.g., Elmegreen et al. 2007). Since our stacking analysis smears out such fine structures, 
we cannot rule out that our LAE sample includes LAEs with irregular morphologies.


\subsection{Relationships between LAEs and LBGs}

Since LAEs and LBGs are two major populations of star-forming galaxies at high redshift,
it has been often discussed what their physical and evolutionary relations are.
LAEs are selected by the narrow-band imaging technique that probes strong Ly$\alpha$ line emission
(see for reviews, Taniguchi 2005, 2008).
This technique does not require that the UV continuum emission is strong enough to be detected
in optical broad-band imaging.
The larger $EW$(Ly$\alpha$) suggest that they 
tend be younger in context of the elapsed time from the onset of star formation activity 
(e.g., Rhoads \& Malhotra 2001; Malhotra \& Rhoads 2002; Nagao et al. 2007).
Therefore, it is expected that LAEs tend to be young less massive star-forming galaxies.
On the other hand, LBGs are selected as
so-called dropout objects in broad-band images (e.g., Steidel et al. 1999). 
This technique requires that the UV continuum is strong enough to be detected. 
However, the selection of dropouts is not affected by the strength of the Ly$\alpha$ line
in principle (e.g., Steidel et al. 2000; Shapley et al. 2003).
Therefore, LBGs tend to be more massive and relatively older than LAEs.

In fact, several studies of LAEs and/or LBGs support the above difference 
between the two populations at $z \sim$ 3 -- 5 as follows.
For example, LAEs tend to have bluer UV continua than LBGs and this property
cannot be explained only with a difference in reddening by dust
(e.g., Gronwall et al. 2007; Ouchi et al. 2008).
However, the number fraction of LAE/LBG
tends to increase with increasing redshift; i.e., from several to 10\% at $z \sim 3$
to $\sim$ 30\% at $z \sim$ 6 -- 7 (e.g., Shimasaku et al. 2006; 
Dow-Hygelund et al. 2007; Ouchi et al. 2008; Sumiya et al. 2009).
Since, at $z \sim$ 6, the elapsed time from the Big Bang is at more 1 Gyr.
LBGs and LAEs may tend to share nearly the same physical properties 
although their Ly$\alpha$ emission luminosities are different on the average. 
Therefore, it seems important to compare observational properties between LBGs and LAEs
at $z \sim 6$ in a more systematic way (e.g., Dow-Hygelund et al. 2007;
Pentericci et al. 2007). Motivated by this, we discuss the size-mass relation 
both for LBGs and LAEs using our own data presented here 
together with all available data from the literature.

Recently, Dow-Hygelund et al. (2007) made a spectroscopic study of 22 LBGs around $z \sim 6$
selected from the samples in three different sky areas and they identified six LBGs
with Ly$\alpha$ emission (i.e., LAEs) at $z = 5.5$ -- 6.1.
They investigated the size-magnitude relation for both LAEs and LBGs compiled from
the literature as well as their own data. They defined the size as the half-light radius 
($R_{\rm HL}$) and the magnitude is ACS $z_{850}$ magnitude (see Fig. 13 in their paper).
They found that the LAEs tend to be more compact than the LBGs at $z \sim 6$,
suggesting that the LAEs are younger than the LBGs. This interpretation may be
supported by other observational differences between LAEs and LBGs;
LAEs tend to be less massive (e.g., Overzier et al. 2006; Gawiser et al. 2006;
Pentericci et al. 2007). 

In order to increase the sample size for an analysis of the size-magnitude relation compared to
Dow-Hygelund et al. (2007), we have also compiled available data from the literature
for $z\sim  6$. Our data compilations is summarized in Table 5.
Since our ACS magnitude is not $z_{850}$ but $I_{814}$, we have to convert our
$I_{814}$ magnitude to $z_{850}$. In this procedure, we assume that the flux of 
the rest-frame continuum at $\lambda <$ 912\AA~ is zero while that at $\lambda \geq$ 912\AA~ is 
constant. We also calculate the contribution of the Ly$\alpha$ flux by using the rest-frame
$EW$(Ly$\alpha$). In this way, we derive the following  equation for conversion;

$z_{850} = I_{814} + 2.5 {\rm log}(0.364 +2.088 \times 10^{-3} EW_0)$.

In Fig. 18, we show our own results together with the compiled data for $z \sim 6$.
Our new data presented in the right panel add more information on this size-magnitude diagram
with respect to that presented by Dow-Hygelund et al. (2007).
We find that there is little difference between the two populations, LAEs and LBGs
at $z \sim 6$ although there many faint LBGs  taken from the very deep ACS imaging of
LBGs made by Bouwens et al. (2006) and Bunker et al. (2003)  are located at  $z_{850} < 29.5$.





\section{CONCLUSIONS}

We have derived the detailed morphological properties of Ly$\alpha$ emitters (LAEs) at $z\approx 5.7$
in the COSMOS field, based on the {\it HST} ACS imaging in the F814W filter.
Our main results and conclusions are summarized below.

(1) Among the 119 LAEs at $z=5.7$ identified in the HST COSMOS field (M07),
85 LAEs are imaged with ACS/F814W. Of those,  47 LAEs are detected 
in our ACS imaging while the remaining 38 ones are not detected. 

(2) All  LAEs detected in ACS have small spatial sizes ($R_{\rm HL} \lesssim 0.4$ arcsec).
However, nearly half of them show a spatially extended morphology with effective radii
larger than 0.15 arcsec (0.93 kpc), which is larger than the PSF size ($R_{\rm HL} =0.11$ arcsec).

(3) Among the 38 ACS not-detected LAEs, 16 LAEs may be spatially extended ($R_{\rm HL} \ge 0.15$ arcsec)
as estimated from their $z'$ and $NB816$ magnitudes.

(4) Comparing the ACS data with our Subaru $NB816$, $i^{\prime}$, and  $z^{\prime}$ data,
we find that the ACS/F814 imaging probes not Ly$\alpha$ line emission but UV continuum arising from
wavelengths longer than  1216 \AA{}.

(5) We find a tendency that LAEs with a larger UV continuum source have a larger Ly$\alpha$ size
as probed by our Subaru NB816 imaging.
Since the LAEs with ACS non-detections have systematically larger $FWHM(NB816)$ than ACS-detected LAEs,
they may have a large UV continuum size and thus their surface brightness in  $I_{814}$ falls below
our detection limit.

(6) UV continuum sizes of LAEs are not directly related to star formation properties such as 
Ly$\alpha$ luminosity and Ly$\alpha$ equivalent widths.
 
\acknowledgements

We would like to thank both the Subaru and HST staff for their invaluable help, and all members of the COSMOS team. 
We would also like to thank the anonymous referee for his/her useful comments.
This work was in part financially supported by JSPS (15340059 and 17253001). 

%

\clearpage


\begin{deluxetable}{lll}
\tablenum{1}
\tabletypesize{\scriptsize}
\tablecaption{SExtractor parameters for source detection}
\tablewidth{0pt}
\tablehead{
\colhead{Parameter} &
\colhead{Value} &
\colhead{Comment}
}

\startdata
DETECT\_TYPE     & CCD & Detector type \\
DETECT\_MINAREA  & 9   & Minimum number of pixels above threshold \\
DETECT\_THRESH   & 1.6 & Detection Threshold in sigma \\
ANALYSIS\_THRESH & 2.0 & Limit for isophotal analysis in sigma \\
FILTER           & Y   & Apply filter for detection \\
FILTER\_NAME     & gauss\_2.0\_5x5.conv & Name of the filter for detection \\
DEBLEND\_NTHRESH & 32  &  Number of deblending sub-thresholds \\
DEBLEND\_MINCONT  & 0.015 & Minimum contrast parameter for deblending \\
CLEAN            & Y     & Clean spurious detection \\
CLEAN\_PARAM      & 1.0   & Cleaning efficiency \\
MASK\_TYPE        & CORRECT & Correct flux for blended source \\
PHOT\_AUTOPARAMS  & 2.5, 0.5  & MAG\_AUTO parameters: Kron factor and minimum radius \\
SATUR\_LEVEL      & 50000     & level (in ADUs) at which arises saturation \\
MAG\_ZEROPOINT    & 25.936    &  Magnitude zero-point \\
MAG\_GAMMA        & 4.0       &  gamma of emulsion (for photographic scans) \\
GAIN             & 1.0       &  Detector gain in e$^{-}$/ADU \\
PIXEL\_SCALE      & 0         &  Size of pixel in arcsec (0=use FITS WCS info) \\
SEEING\_FWHM      & 0.11      &  Stellar FWHM in arcsec \\
BACK\_SIZE        & 64        &  Background mesh size \\
BACK\_FILTERSIZE  & 3         &  Background filter size \\
BACKPHOTO\_TYPE   & GLOBAL    &  Photometry background subtraction type 
\enddata

\end{deluxetable}

\begin{deluxetable}{lcccc}
\tablenum{2}
\tablecaption{COSMOS $z \approx 5.7$ LAE  sample}
\tablewidth{0pt}
\tablehead{
\colhead{LAE sample} &
\colhead{Number of LAEs} &
\colhead{Spectroscopic confirmation}
}

\startdata
Total (Murayama 2007)                        & 119 & 24 \\
\hspace{1em}Out of the ACS/F814W field & 34  & 4 \\
\hspace{1em}In the ACS/F814W field & 85  & 20 \\
\hspace{2em}ACS/F814W not-detected   & 38  & 3 \\
\hspace{2em}ACS/F814W detected     & 47  & 17 \\
\hspace{4em}Double                 &  2  &  2 \\
\enddata

\end{deluxetable}

\begin{deluxetable}{ccccc}
\tablenum{3}
\tablecaption{ACS F814W Properties for the LAEs at $z \approx 5.7$}
\tablewidth{0pt}
\tablehead{
\colhead{Number\tablenotemark{a}} &
\colhead{$I_{814}$\tablenotemark{b}} &
\colhead{$R_{\rm HL}$\tablenotemark{c}} &
\colhead{$I_{814}$ (estimated) \tablenotemark{d}}&
\colhead{$R_{\rm HL}$  (estimated) \tablenotemark{d}} \\
\colhead{} &
\colhead{(mag)} &
\colhead{(arcsec)} &
\colhead{(mag)} &
\colhead{(arcsec)}
}

\startdata
\multicolumn{5}{c}{Detected} \\  
\hline
2     &        26.9     $\pm$    0.1     &    0.11     & $26.8_{-0.3}^{+0.4}$ & $0.10_{-0.05}^{+0.10}$ \\
4     &  26.5  $\pm$ 0.1  & 0.20                       & $26.1_{-0.7}^{+0.7}$ & $0.35_{-0.18}^{+0.30}$ \\
5     &        26.7     $\pm$    0.1     &    0.13     & $26.6_{-0.4}^{+0.4}$ & $0.13_{-0.05}^{+0.10}$ \\
7     &        27.5     $\pm$    0.2     &    0.10     & $27.3_{-0.6}^{+0.4}$ & $0.08_{-0.05}^{+0.18}$ \\
13    &        26.8     $\pm$    0.1     &    0.09     & $26.7_{-0.2}^{+0.3}$ & $0.08_{-0.05}^{+0.08}$ \\
14    &  25.7  $\pm$ 0.1  & 0.24                       & $25.4_{-0.4}^{+0.5}$ & $0.38_{-0.15}^{+0.28}$ \\
15    &  26.0  $\pm$ 0.1  & 0.24                       & $25.9_{-0.4}^{+0.5}$ & $0.25_{-0.10}^{+0.20}$ \\
20    &  26.4  $\pm$ 0.1  & 0.23                       & $26.1_{-0.7}^{+0.8}$ & $0.38_{-0.20}^{+0.35}$ \\
23    &        27.1     $\pm$    0.2     &    0.13     & $27.0_{-0.6}^{+0.6}$ & $0.13_{-0.08}^{+0.23}$ \\
30    &        27.0     $\pm$    0.1     &    0.07     & $26.9_{-0.3}^{+0.3}$ & $0.08_{-0.05}^{+0.08}$ \\
34    &  26.4  $\pm$ 0.1  & 0.25                       & $26.2_{-0.5}^{+0.5}$ & $0.28_{-0.13}^{+0.20}$ \\
35    &  26.3  $\pm$ 0.10  & 0.20                      & $26.2_{-0.3}^{+0.3}$ & $0.18_{-0.08}^{+0.13}$ \\
39    &        26.8     $\pm$    0.1     &    0.10     & $26.7_{-0.3}^{+0.4}$ & $0.10_{-0.08}^{+0.08}$ \\
40    &  25.4  $\pm$ 0.1  & 0.36                       & $25.3_{-0.3}^{+0.6}$ & $0.43_{-0.20}^{+0.18}$ \\
41    &  26.3  $\pm$ 0.1  & 0.19                       & $26.3_{-0.5}^{+0.4}$ & $0.18_{-0.08}^{+0.15}$ \\
42    &        27.7     $\pm$    0.1     &    0.08     & $27.5_{-0.3}^{+0.4}$ & $0.05_{-0.03}^{+0.05}$ \\
43    &        27.3     $\pm$    0.2     &    0.12     & $27.0_{-0.9}^{+0.9}$ & $0.33_{-0.20}^{+0.40}$ \\
44    &        27.1     $\pm$    0.2     &    0.14     & $26.9_{-0.8}^{+0.6}$ & $0.18_{-0.10}^{+0.33}$ \\
45    &  26.6  $\pm$ 0.1  & 0.18                       & $26.4_{-0.8}^{+0.6}$ & $0.28_{-0.15}^{+0.33}$ \\
47    &  26.1  $\pm$ 0.1  & 0.16                       & $26.0_{-0.2}^{+0.3}$ & $0.15_{-0.05}^{+0.08}$ \\
49    &  26.5  $\pm$ 0.2  & 0.20                       & $26.0_{-0.8}^{+0.7}$ & $0.38_{-0.18}^{+0.33}$ \\
50    &        26.4     $\pm$    0.1     &    0.14     & $26.4_{-0.3}^{+0.3}$ & $0.13_{-0.08}^{+0.08}$ \\
51    &        26.9     $\pm$    0.2     &    0.11     & $26.9_{-0.5}^{+0.5}$ & $0.13_{-0.08}^{+0.18}$ \\
55    &  27.1  $\pm$ 0.2  & 0.15                       & $26.5_{-0.9}^{+1.0}$ & $0.25_{-0.13}^{+0.35}$ \\
60\tablenotemark{e}  &  26.8\tablenotemark{f}  &  0.94\tablenotemark{g}   &  \nodata & \nodata \\
60a   &  26.9  $\pm$ 0.2  & 0.15                      & $26.7_{-0.8}^{+0.7}$ & $0.23_{-0.13}^{+0.35}$ \\
60b   &  27.3  $\pm$ 0.2  & 0.11                      & $27.1_{-0.6}^{+0.5}$ & $0.10_{-0.05}^{+0.20}$  \\
68    &  25.9  $\pm$ 0.1  & 0.40                      & $25.7_{-0.6}^{+0.5}$ & $0.35_{-0.13}^{+0.33}$ \\
69    &        27.8     $\pm$    0.2     &    0.08    & $27.5_{-0.4}^{+0.1}$ & $0.03_{-0.03}^{+0.13}$ \\
71    &        26.1     $\pm$    0.1     &    0.10    & $26.1_{-0.1}^{+0.2}$ & $0.08_{-0.05}^{+0.05}$ \\
73    &        26.8     $\pm$    0.1     &    0.13    & $26.7_{-0.4}^{+0.4}$ & $0.13_{-0.05}^{+0.13}$ \\
74    &  26.9  $\pm$ 0.2  & 0.16                      & $26.6_{-1.2}^{+1.2}$ & $0.35_{-0.23}^{+0.40}$ \\
75    &  26.0  $\pm$ 0.1  & 0.19                      & $26.0_{-0.3}^{+0.3}$ & $0.15_{-0.05}^{+0.10}$ \\
76    &        26.7     $\pm$    0.1     &    0.10    & $26.7_{-0.3}^{+0.3}$ & $0.08_{-0.05}^{+0.08}$ \\
77    &        27.7     $\pm$    0.1     &    0.07    & $27.5_{-0.3}^{+0.4}$ & $0.05_{-0.03}^{+0.05}$ \\
78    &  25.7  $\pm$ 0.1  & 0.30                      & $25.4_{-0.4}^{+0.5}$ & $0.35_{-0.13}^{+0.25}$ \\
81    &  27.1  $\pm$ 0.1  & 0.16                      & $26.9_{-0.9}^{+0.7}$ & $0.28_{-0.18}^{+0.43}$ \\
83    &        27.3     $\pm$    0.1     &    0.09    & $27.2_{-0.4}^{+0.4}$ & $0.08_{-0.05}^{+0.13}$ \\
84    &        26.2     $\pm$    0.1     &    0.15    & $26.2_{-0.3}^{+0.3}$ & $0.13_{-0.05}^{+0.08}$ \\
85    &  26.5  $\pm$ 0.1  & 0.17                      & $26.4_{-0.3}^{+0.3}$ & $0.15_{-0.05}^{+0.10}$ \\
86    &        27.6     $\pm$    0.2     &    0.12    & $27.3_{-0.9}^{+0.4}$ & $0.10_{-0.05}^{+0.28}$ \\
95    &        27.0     $\pm$    0.1     &    0.11    & $26.9_{-0.4}^{+0.4}$ & $0.10_{-0.05}^{+0.10}$ \\
96    &  26.6  $\pm$ 0.1  & 0.19                      & $26.4_{-0.7}^{+0.6}$ & $0.25_{-0.13}^{+0.30}$ \\
98    &        26.0     $\pm$    0.1     &    0.14    & $26.0_{-0.2}^{+0.3}$ & $0.10_{-0.05}^{+0.08}$ \\
99    &  26.8  $\pm$ 0.1  & 0.16                      & $26.6_{-0.4}^{+0.5}$ & $0.18_{-0.08}^{+0.10}$ \\
104    &  26.5  $\pm$ 0.2  & 0.20                     & $26.3_{-0.7}^{+0.7}$ & $0.30_{-0.15}^{+0.35}$ \\
105    &        25.6     $\pm$    0.1     &    0.10   & $25.6_{-0.1}^{+0.1}$ & $0.08_{-0.05}^{+0.03}$ \\
107    &        27.0     $\pm$    0.2     &    0.13   & $26.8_{-0.9}^{+0.7}$ & $0.20_{-0.13}^{+0.35}$ \\
110\tablenotemark{e}  &  26.4\tablenotemark{f}  &  0.36\tablenotemark{g}   &  \nodata  & \nodata \\
110a   &  26.5  $\pm$ 0.1  & 0.17                     & $26.5_{-0.5}^{+0.4}$ & $0.18_{-0.10}^{+0.13}$ \\
110b   &  27.1  $\pm$ 0.2  & 0.18                     & $26.8_{-0.9}^{+0.7}$ & $0.28_{-0.15}^{+0.35}$ \\
\hline
\multicolumn{5}{c}{Not-Detected} \\  
\hline
1 & $>$ 27.4   & \nodata & \nodata & \nodata \\
6 & $>$ 27.1   & \nodata & \nodata & \nodata \\
9 & $>$ 27.3   & \nodata & \nodata & \nodata \\
16 & $>$ 27.4   & \nodata & \nodata & \nodata \\
19 & $>$ 27.2   & \nodata & \nodata & \nodata \\
21 & $>$ 27.4   & \nodata & \nodata & \nodata \\
24 & $>$ 27.4   & \nodata & \nodata & \nodata \\
25 & $>$ 27.5   & \nodata & \nodata & \nodata \\
26 & $>$ 27.4   & \nodata & \nodata & \nodata \\
27 & $>$ 27.3   & \nodata & \nodata & \nodata \\
28 & $>$ 27.3   & \nodata & \nodata & \nodata \\
29 & $>$ 27.4   & \nodata & \nodata & \nodata \\
38 & $>$ 27.3   & \nodata & \nodata & \nodata \\
48 & $>$ 27.5   & \nodata & \nodata & \nodata \\
54 & $>$ 27.5   & \nodata & \nodata & \nodata \\
57 & $>$ 27.1   & \nodata & \nodata & \nodata \\
58 & $>$ 27.4   & \nodata & \nodata & \nodata \\
59 & $>$ 27.8   & \nodata & \nodata & \nodata \\
61 & $>$ 27.5   & \nodata & \nodata & \nodata \\
62 & $>$ 27.4   & \nodata & \nodata & \nodata \\
79 & $>$ 27.6   & \nodata & \nodata & \nodata \\
80 & $>$ 26.9   & \nodata & \nodata & \nodata \\
87 & $>$ 27.2   & \nodata & \nodata & \nodata \\
88 & $>$ 27.5   & \nodata & \nodata & \nodata \\
89 & $>$ 27.3   & \nodata & \nodata & \nodata \\
91 & $>$ 27.4   & \nodata & \nodata & \nodata \\
92 & $>$ 27.6   & \nodata & \nodata & \nodata \\
93 & $>$ 27.3   & \nodata & \nodata & \nodata \\
97 & $>$ 27.2   & \nodata & \nodata & \nodata \\
100 & $>$ 27.3   & \nodata & \nodata & \nodata \\
101 & $>$ 27.6   & \nodata & \nodata & \nodata \\
102 & $>$ 27.3   & \nodata & \nodata & \nodata \\
103 & $>$ 26.9   & \nodata & \nodata & \nodata \\
106 & $>$ 26.7   & \nodata & \nodata & \nodata \\
108 & $>$ 27.4   & \nodata & \nodata & \nodata \\
111 & $>$ 27.2   & \nodata & \nodata & \nodata \\
112 & $>$ 27.3   & \nodata & \nodata & \nodata \\
114 & $>$ 27.1   & \nodata & \nodata & \nodata
\enddata

\tablenotetext{a}{The LAE IDs given in Murayama et al.~(2007).}
\tablenotetext{b}{AB magnitude. Lower-limits represent 3$\sigma$ significance with assuming a 1-arcsec diameter aperture.}
\tablenotetext{c}{Effective radius.}
\tablenotetext{d}{Estimated values by Monte Carlo calculations.}
\tablenotetext{e}{Double ACS sources.}
\tablenotetext{f}{Magnitude corresponding to the sum of flux of each source.}
\tablenotetext{g}{Separation between the double sources.}
\end{deluxetable}

\clearpage


\begin{deluxetable}{cccccccccc}
\rotate
\tablenum{4}
\tabletypesize{\small}
\tablecaption{Photometric properties from Subaru Suprime-Cam Images for the LAEs at $z \approx 5.7$}
\tablewidth{0pt}
\tablehead{
\colhead{Number\tablenotemark{a}} &
\colhead{$i'$\tablenotemark{b}} &
\colhead{$z'$\tablenotemark{b}} &
\colhead{$NB816$\tablenotemark{b}} &
\colhead{$FWHM$(NB816)\tablenotemark{c}} &
\colhead{$\epsilon$(NB816)\tablenotemark{d}} &
\colhead{$PA$(NB816)} &
\colhead{$L({\rm Ly}\alpha)$} &
\colhead{$EW_0({\rm Ly}\alpha)$\tablenotemark{e}} &
\colhead{$z_{\rm sp}$\tablenotemark{f}} \\
\colhead{} &
\colhead{($\phi$3\farcs 0)} &
\colhead{($\phi$3\farcs 0)} &
\colhead{($\phi$3\farcs 0)} &
\colhead{(arcsec)} &
\colhead{} &
\colhead{(degree)} &
\colhead{($10^{42}$ ergs s$^{-1}$)} &
\colhead{(\AA)} &
\colhead{}
}

\startdata
\multicolumn{10}{c}{Detected} \\  
\hline
2 & 26.5  & 25.9  & 24.6  & 1.1  & 0.17  & 9  & 8.0  &  46 & \nodata \\
4 & 24.5  & 24.8  & 23.2  & 1.4  & 0.08  & 168  & 31.0  &  67  & \nodata \\
5 & 27.1  & 26.2  & 24.4  & 1.2  & 0.04  & 141  & 10.7  &  86  & \nodata \\
7 & 26.0  & 25.6  & 24.6  & 1.3  & 0.11  & 179  & 7.9  &  34  & \nodata \\
13 & 27.6  & 99.0  & 23.9  & 1.1  & 0.07  & 68  & 18.4  & $>$ 175 & \nodata  \\
14 & 24.9  & 25.0  & 23.6  & 1.4  & 0.15  & 74  & 21.2  &  55  & \nodata \\
15 & 26.2  & 25.1  & 24.6  & 1.0  & 0.06  & 92  & 6.7  &  19  & \nodata \\
20 & 26.5  & 26.0  & 24.7  & 1.0  & 0.05  & 106  & 8.0  &  51  & \nodata \\
23 & 25.4  & 25.9  & 24.4  & 1.4  & 0.37  & 91  & 10.1  &  58  & \nodata \\
30 & 25.9  & 25.8  & 24.3  & 1.0  & 0.11  & 127  & 12.2  &  66  & \nodata \\
34 & 26.4  & 25.5  & 24.7  & 1.7  & 0.31  & 8  & 7.3  &  29  & 5.681\\
35 & 25.5  & 25.6  & 24.4  & 1.5  & 0.41  & 14  & 9.6  &  42 & 5.663 \\
39 & 25.9  & 26.0  & 24.2  & 1.1  & 0.16  & 154  & 13.6  &  89 & 5.718 \\
40 & 25.5  & 24.4  & 23.9  & 1.5  & 0.09  & 174  & 13.3  &  20 & 5.690 \\
41 & 27.1  & 99.0  & 24.5  & 1.0  & 0.16  & 6  & 11.2  & $>$ 103  & \nodata \\
42 & 27.0  & 27.3  & 24.8  & 1.1  & 0.06  & 137  & 8.0  & $>$ 72  &5.681 \\
43 & 26.3  & 26.1  & 24.2  & 1.3  & 0.10  & 166  & 13.2  &  91 &5.657 \\
44 & 26.3  & 26.0  & 24.3  & 1.4  & 0.31  & 67  & 11.6  &  77  &  5.711 \\
45 & 25.7  & 25.9  & 24.5  & 1.4  & 0.20  & 97  & 9.6  &  55  &  5.668 \\
47 & 25.5  & 25.2  & 24.1  & 1.4  & 0.18  & 128  & 13.5  &  42 &  5.714 \\
49 & 26.8  & 28.3  & 24.8  & 1.1  & 0.24  & 176  & 8.0  & $>$ 84  & \nodata \\
50 & 26.0  & 99.0  & 23.5  & 1.2  & 0.06  & 160  & 27.9  & $>$ 260  & \nodata \\
51 & 27.0  & 99.0  & 24.9  & 1.2  & 0.26  & 22  & 9.5  & $>$ 88  & \nodata \\
55 & 27.3  & 26.6  & 24.5  & 1.3  & 0.19  & 139  & 9.8  & $>$ 97  & \nodata \\
60 & 25.4  & 25.5  & 24.1  & 1.5  & 0.22  & 79  & 13.7  &  56   & 5.661 \\
68 & 26.4  & 26.7  & 24.5  & 1.0  & 0.06  & 75  & 10.4  & $>$ 111 & 5.660  \\
69 & 27.8  & 99.0  & 24.6  & 1.3  & 0.10  & 49  & 10.0  & $>$ 101  & \nodata \\
71 & 25.4  & 25.3  & 23.3  & 1.1  & 0.05  & 50  & 31.1  &  105  & \nodata \\
73 & 25.2  & 25.7  & 24.0  & 1.1  & 0.09  & 14  & 15.0  &  73 & 5.679   \\
74 & 26.2  & 25.6  & 24.4  & 1.9  & 0.31  & 53  & 9.9  &  44  & \nodata \\
75 & 26.7  & 26.0  & 24.2  & 1.2  & 0.07  & 147  & 13.4  &  90 &5.721   \\
76 & 29.7  & 25.6  & 24.0  & 1.1  & 0.06  & 14  & 14.8  &  64 &5.678  \\
77 & 26.2  & 25.6  & 24.3  & 1.2  & 0.10  & 62  & 11.5  &  53 &5.782   \\
78 & 25.7  & 25.0  & 23.5  & 1.4  & 0.01  & 164  & 23.8  &  64  & \nodata \\
81 & 25.7  & 26.4  & 24.6  & 1.5  & 0.36  & 116  & 8.9  & $>$ 76  & \nodata \\
83 & 26.0  & 25.6  & 24.7  & 1.2  & 0.19  & 58  & 7.1  &  31  & \nodata \\
84 & 24.6  & 24.0  & 23.5  & 1.5  & 0.22  & 34  & 18.1  &  19   & \nodata\\
85 & 25.3  & 24.5  & 23.8  & 1.5  & 0.17  & 19  & 16.4  &  27  & \nodata \\
86 & 26.7  & 26.3  & 24.8  & 1.2  & 0.20  & 154  & 7.7  &  64  & \nodata \\
95 & 26.7  & 26.9  & 24.7  & 1.1  & 0.22  & 14  & 8.5  & $>$ 79   & \nodata\\
96 & 25.9  & 25.5  & 24.1  & 1.1  & 0.16  & 157  & 13.5  &  56 & 5.673 \\
98 & 26.1  & 25.5  & 24.0  & 1.1  & 0.06  & 78  & 14.8  &  58  & \nodata \\
99 & 25.8  & 26.1  & 24.6  & 1.3  & 0.20  & 166  & 8.7  &  64  & \nodata \\
104 & 25.8  & 26.2  & 23.7  & 1.0  & 0.07  & 22  & 22.3  &  170  & \nodata \\
105 & 99.0  & 99.0  & 25.1  & 1.1  & 0.24  & 171  & 6.5  & $>$ 59  & \nodata \\
107 & 26.5  & 27.7  & 24.6  & 1.3  & 0.20  & 162  & 10.1  & $>$ 92  & \nodata \\
110 & 26.2  & 24.7  & 24.3  & 1.2  & 0.05  & 48  & 8.4  &  16 & 5.672  \\
\hline
\multicolumn{10}{c}{Not-Detected} \\  
\hline
1 & 25.8  & 25.9  & 24.5  & 1.7  & 0.31  & 71  & 9.2  &  53  & \nodata \\
6 & 27.1  & 26.4  & 24.7  & 1.2  & 0.16  & 97  & 8.1  & $>$ 74  & \nodata \\
9 & 27.3  & 30.4  & 24.6  & 1.2  & 0.31  & 168  & 10.4  & $>$ 98  & \nodata \\
16 & 26.9  & 27.5  & 24.9  & 1.1  & 0.21  & 84  & 7.1  & $>$ 68  & \nodata \\
19 & 25.2  & 25.9  & 24.2  & 1.4  & 0.31  & 136  & 13.0  &  77   & \nodata\\
21 & 26.7  & 27.5  & 24.8  & 1.2  & 0.25  & 136  & 8.4  & $>$ 79  & \nodata \\
24 & 26.4  & 26.4  & 24.3  & 1.2  & 0.03  & 97  & 12.3  & $>$ 100  & \nodata \\
25 & 26.3  & 26.7  & 24.6  & 1.5  & 0.10  & 115  & 9.3  & $>$ 94  & \nodata \\
26 & 26.3  & 28.7  & 25.1  & 0.9  & 0.04  & 49  & 6.5  & $>$ 71  & \nodata \\
27 & 25.8  & 25.8  & 24.0  & 2.7  & 0.16  & 180  & 16.3  &  85  & \nodata \\
28 & 26.1  & 27.0  & 24.8  & 1.4  & 0.12  & 19  & 7.6  & $>$ 77  & \nodata \\
29 & 25.5  & 26.0  & 24.4  & 2.0  & 0.29  & 170  & 11.1  &  72  & \nodata \\
38 & 26.7  & 99.0  & 24.5  & 1.0  & 0.05  & 81  & 10.9  & $>$ 100  & \nodata \\
48 & 27.3  & 99.0  & 24.2  & 1.4  & 0.15  & 36  & 14.8  & $>$ 124  & \nodata \\
54 & 25.0  & 26.9  & 24.5  & 1.2  & 0.31  & 150  & 10.4  & $>$ 97  & \nodata \\
57 & 25.1  & 24.9  & 23.5  & 1.2  & 0.15  & 7  & 22.8  &  51  & \nodata \\
58 & 28.2  & 99.0  & 24.8  & 1.0  & 0.10  & 46  & 8.5  & $>$ 78  & \nodata \\
59 & 27.0  & 26.3  & 24.7  & 1.1  & 0.13  & 145  & 8.3  &  68  & \nodata \\
61 & 25.3  & 24.3  & 24.0  & 1.4  & 0.17  & 164  & 10.7  &  15  & \nodata \\
62 & 27.0  & 99.0  & 24.5  & 1.2  & 0.08  & 133  & 11.5  & $>$ 102  & \nodata \\
79 & 27.4  & 26.1  & 24.4  & 1.2  & 0.16  & 65  & 11.2  &  79  & 5.686 \\
80 & 25.7  & 25.5  & 24.3  & 0.9  & 0.06  & 57  & 10.6  &  44  & \nodata \\
87 & 26.2  & 25.9  & 24.7  & 1.4  & 0.30  & 146  & 7.8  &  46  & \nodata \\
88 & 27.3  & 99.0  & 24.5  & 1.4  & 0.15  & 38  & 11.4  & $>$ 105  & \nodata \\
89 & 26.2  & 29.6  & 24.5  & 1.4  & 0.38  & 173  & 10.8  & $>$ 92  & \nodata \\
91 & 25.5  & 29.7  & 24.6  & 1.4  & 0.13  & 166  & 9.9  & $>$ 89   & \nodata\\
92 & 27.0  & 99.0  & 24.7  & 1.3  & 0.10  & 133  & 9.5  & $>$ 84  & \nodata \\
93 & 28.0  & 99.0  & 24.9  & 1.5  & 0.28  & 61  & 8.8  & $>$ 79  & \nodata \\
97 & 26.2  & 25.6  & 24.4  & 1.4  & 0.29  & 34  & 10.2  &  46  & \nodata \\
100 & 26.6  & 26.3  & 24.8  & 1.4  & 0.24  & 13  & 7.1  &  58  & \nodata \\
101 & 27.1  & 99.0  & 24.8  & 1.3  & 0.34  & 59  & 10.2  & $>$ 98   & \nodata\\
102 & 26.6  & 26.0  & 24.7  & 0.9  & 0.19  & 0  & 7.6  &  50  & \nodata \\
103 & 28.8  & 26.1  & 24.4  & 1.4  & 0.23  & 171  & 11.2  &  83  & \nodata \\
106 & 27.1  & 99.0  & 24.7  & 1.6  & 0.20  & 98  & 10.2  & $>$ 100  & \nodata \\
108 & 26.0  & 26.1  & 24.2  & 0.9  & 0.15  & 66  & 13.4  &  93 &5.693  \\
111 & 25.4  & 24.3  & 23.9  & 1.8  & 0.08  & 45  & 12.6  &  17  & \nodata \\
112 & 26.7  & 26.0  & 24.6  & 1.5  & 0.03  & 91  & 8.8  &  59  & \nodata \\
114 & 26.4  & 99.0  & 24.0  & 1.6  & 0.25  & 11  & 18.1  & $>$ 155 & 5.627  \\
\hline
\multicolumn{10}{c}{No ACS Image} \\  
\hline
3&26.4 &26.3 &24.7 &1.7 &0.42 &165 &8.2 &68  & \nodata\\
8&25.7 &24.8 &24.2 &2.0 &0.33 &84 &9.8 &20  & \nodata\\
10&25.4 &24.6 &24.1 &1.9 &0.27 &130 &11.8 &22  & \nodata\\
11&24.9 &24.8 &24.1 &2.3 &0.24 &107 &12.2 &25 & \nodata \\
12&28.1 &99.0 &24.7 &1.0 &0.13 &62 &8.8 &$>$69 & \nodata \\
17&25.4 &25.2 &24.2 &1.3 &0.25 &38 &11.7 &35  & \nodata\\
18&25.3 &25.7 &23.8 &1.2 &0.11 &43 &19.1 &91  & \nodata\\
22&26.7 &99.0 &24.3 &1.9 &0.24 &119 &12.8 &$>$74 & \nodata \\
31&25.8 &26.2 &23.8 &1.4 &0.14 &36 &20.3 &154 & \nodata \\
32&26.6 &26.1 &23.9 &1.3 &0.17 &55 &18.5 &127  & \nodata\\
33&25.8 &25.1 &24.2 &1.1 &0.08 &74 &11.0 &30 & 5.639  \\
36&27.2 &25.9 &24.5 &1.5 &0.15 &39 &9.4 &56 & \nodata \\
37&26.7 &99.0 &24.4 &1.0 &0.02 &50 &12.4 &$>$100 & \nodata \\
46&24.8 &24.9 &23.8 &1.3 &0.20 &137 &16.7 &39 & \nodata \\
52&25.8 &25.0 &24.2 &1.6 &0.31 &34 &10.7 &29 & \nodata \\
53&25.1 &24.9 &23.5 &1.1 &0.10 &63 &22.8 &53 & \nodata \\
56&26.3 &25.9 &24.6 &1.6 &0.28 &8 &8.2 &47  & \nodata\\
63&26.0 &25.5 &24.2 &1.4 &0.14 &80 &11.8 &48  & \nodata\\
64&27.1 &25.8 &24.5 &1.5 &0.11 &95 &9.4 &49 &5.680  \\
65&26.3 &26.4 &24.8 &1.0 &0.16 &5 &7.4 &68 & 5.693 \\
66&26.7 &24.5 &24.3 &1.1 &0.10 &89 &7.1 &11 & 5.656  \\
67&24.7 &23.7 &23.3 &1.6 &0.12 &86 &20.8 &15 & \nodata \\
70&26.1 &27.5 &24.3 &1.0 &0.08 &14 &12.6 &$>$89 & \nodata \\
72&26.6 &25.8 &24.8 &1.1 &0.12 &49 &6.5 &35  & \nodata\\
82&26.0 &25.4 &24.6 &1.2 &0.18 &48 &7.7 &30 & \nodata \\
90&26.5 &99.0 &24.7 &1.4 &0.15 &163 &9.0 &$>$78 & \nodata \\
94&26.3 &27.1 &24.8 &1.3 &0.30 &4 &8.0 &$>$64 & \nodata \\
109&25.9 &25.3 &24.3 &1.1 &0.26 &44 &10.9 &37 & \nodata \\
113&25.6 &99.0 &24.6 &0.9 &0.15 &146 &10.5 &$>$92  & \nodata\\
115&24.7 &24.6 &23.4 &1.1 &0.10 &18 &26.0 &45 & \nodata \\
116&26.3 &25.4 &24.7 &1.0 &0.02 &27 &6.3 &23 & \nodata \\
117&25.1 &24.8 &24.1 &1.7 &0.17 &163 &11.4 &25 & \nodata \\
118&25.6 &25.5 &23.6 &1.3 &0.17 &53 &22.8 &94 & \nodata \\
119&25.8 &25.2 &24.5 &1.7 &0.36 &41 &8.2 &25 & \nodata
\enddata

\tablenotetext{a}{The LAE IDs given in Murayama et al.~(2007).}
\tablenotetext{b}{AB magnitude.
 An entry of ``99.0'' indicates that no excess flux was measured.}
\tablenotetext{c}{Spatial sizes measured on NB816 images.}
\tablenotetext{d}{Ellipticity measured on NB816 images.}
\tablenotetext{e}{\ Lower-limits represent 1$\sigma$ significance.}
\tablenotetext{f}{Spectroscopic redshift.}

\end{deluxetable}

\clearpage

\begin{deluxetable}{cp{8cm}l}
\tablenum{5}
\tabletypesize{\small}
\tablecaption{Symbols and references for Figure 18}
\tablewidth{0pt}
\tablehead{
\colhead{Symbol} &
\colhead{Description} &
\colhead{References} 
}
\startdata
Large red filled circle&
COSMOS LAEs at $z\approx 5.7$ &
this study \\
Small black filled circle&
$i$ band dropout galaxies in UDF, UDF-P, GOODS-N, and GOODS-S &
Bouwens et al. (2006) \\
Small red filled circle&
Spectroscopic confirmed LAEs among $i$ band dropout galaxies in  GOODS-S &
Bouwens et al. (2006)\\
 & &  ESO GOODS-S database\tablenotemark{a}\\
Small blue filled circle&
Spectroscopic confirmed LAEs among $i$ band dropout galaxies in GOODS-S &
Bouwens et al. (2006) \\
 & &  ESO GOODS-S database\tablenotemark{a}\\
Red plus &
LAE at $z=5.7$ &
Bunker et al. (2003) \\
Black asterisk &
$i$ band dropout galaxies in UDF &
Bunker et al. (2004) \\
Red open circle&
LAEs at $z\sim 5.8$ &
Stanway et al. (2004a) \\
Red cross&
LAEs in GOODS-N &
Stanway et al. (2004b) \\
Blue open square &
Non emission-line galaxies at $z\sim 6$ &
Dow-Hygelund et al. (2007) \\
Red open square &
Spectroscopic confirmed LAEs at $z\sim 6$ &
Dow-Hygelund et al. (2007) \\
\enddata

\tablenotetext{a}{Spectroscopic properties for the sample from Bouwens et al. (2006)
were taken from the ESO GOODS/VIRMOS spectroscopic database at http://www.eso.org/science/goods/.
}

\end{deluxetable}


\begin{figure}
\plotone{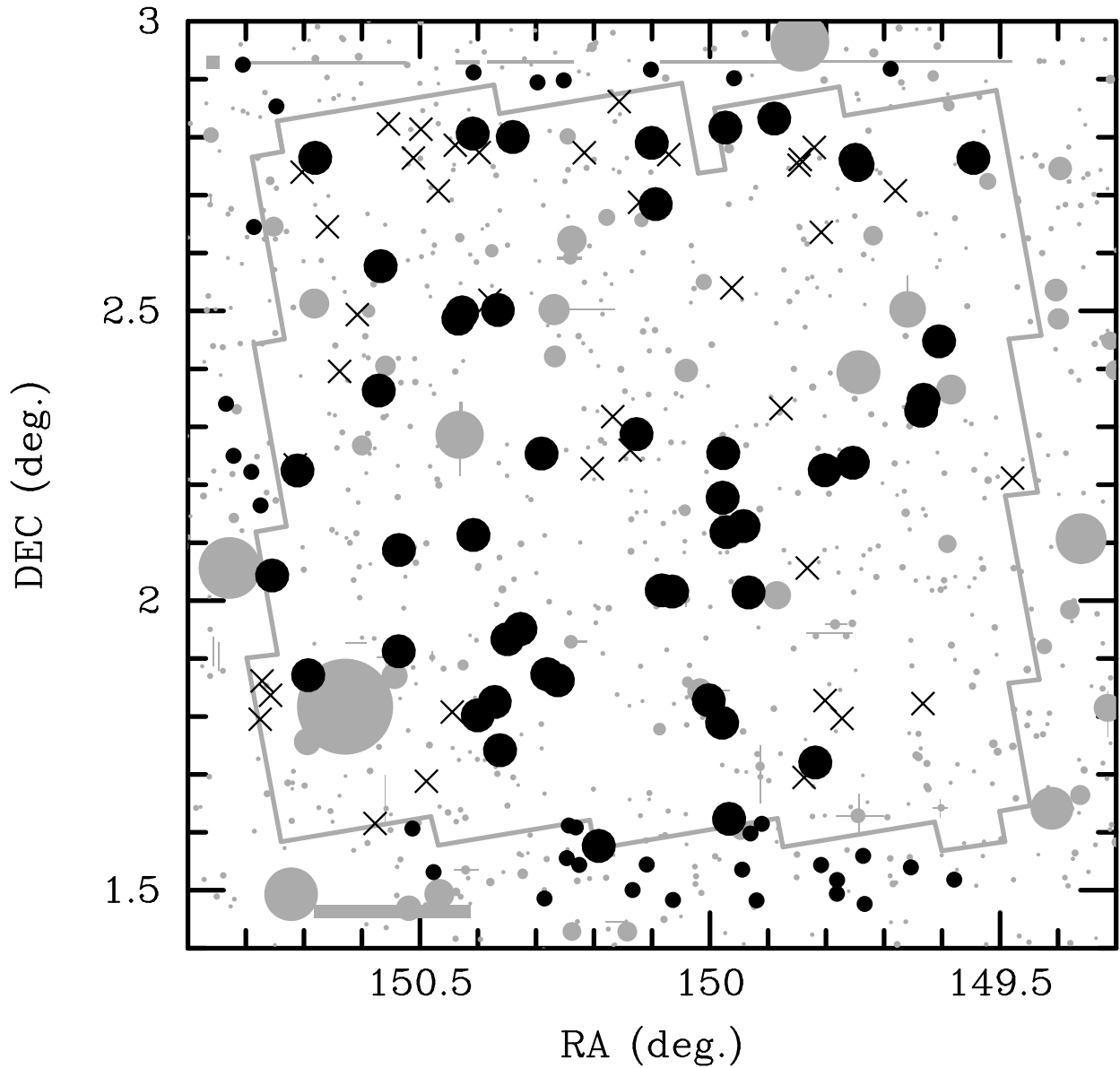}
\caption{Spatial distribution of our sample of 119 LAEs (M07).
The whole area is 1.95 square degree that is mapped with our Suprime-Cam observations.
The ACS image is available for 1.64 square degree, its footprint is indicated by the solid gray line. 
Masked out regions are shown by filled gray circles or thin gray lines.
The 47 LAEs detected with ACS are shown
by large filled  circles while 38 LAEs undetected in the ACS images are shown by  crosses.
The remaining 34 LAEs shown by dots fall outside the HST/ACS field.}
\end{figure}
\clearpage

\begin{figure}
\epsscale{0.8}
\plottwo{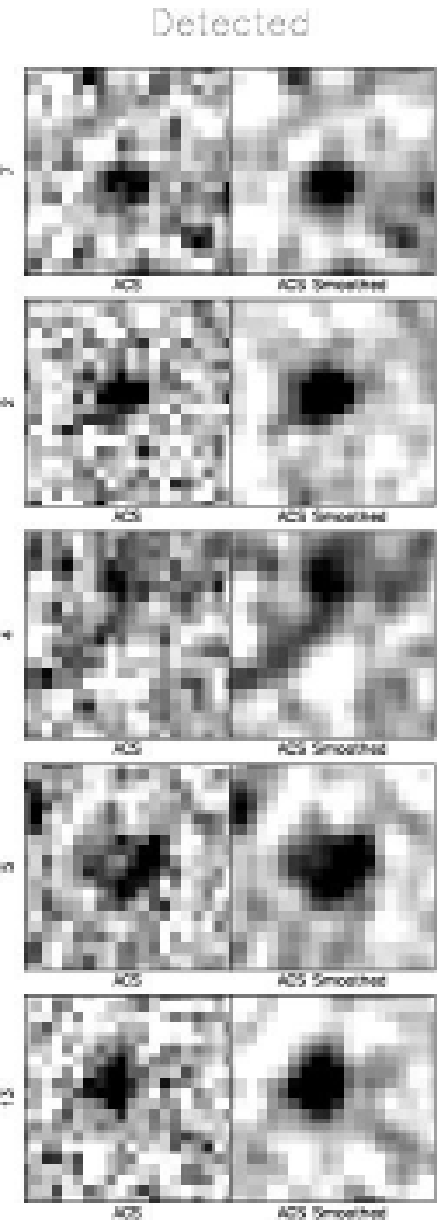}{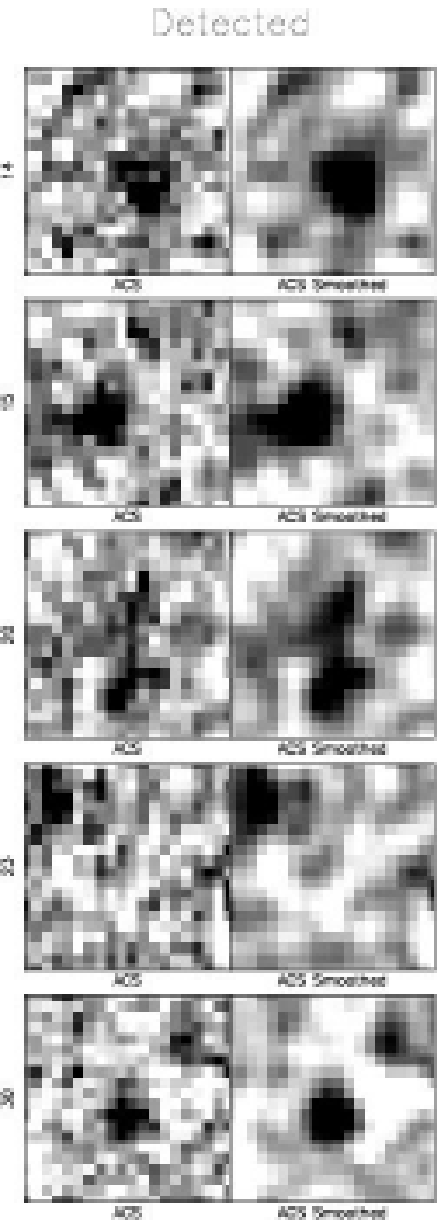}
\caption{Thumbnails of all  85 LAEs analyzed in this paper. 
Each panel has a size of 1$^{\prime\prime} \times$
1$^{\prime\prime}$. North is up and east is left.}
\end{figure}
\clearpage

\plottwo{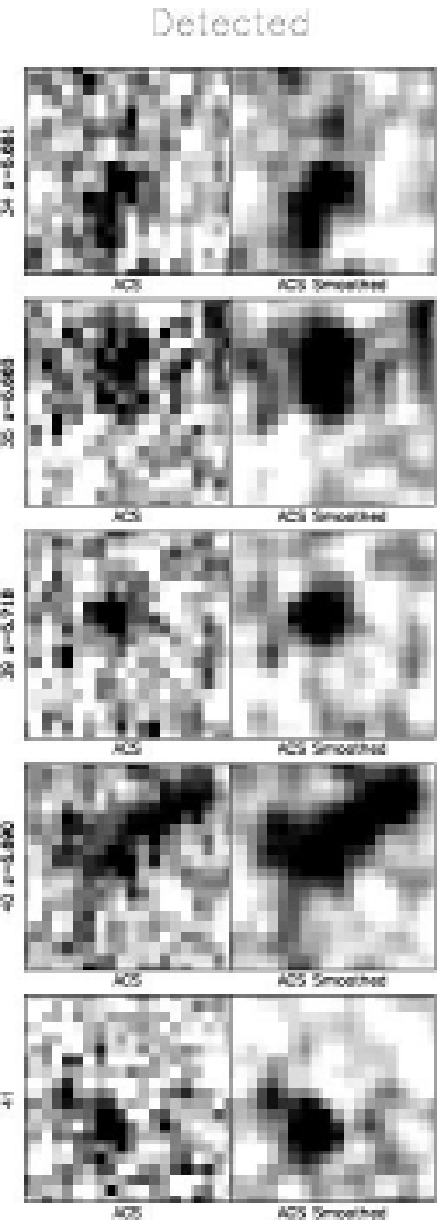}{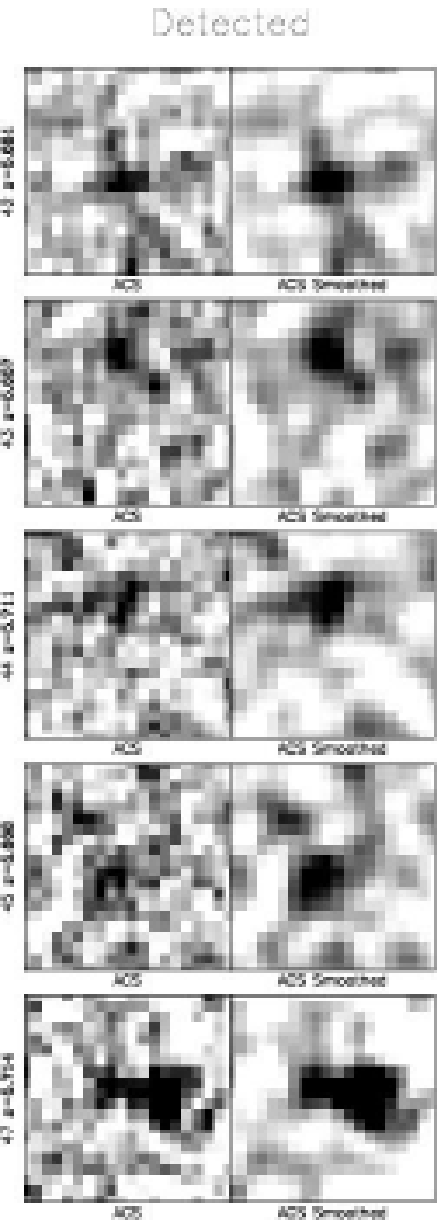}
\clearpage

\plottwo{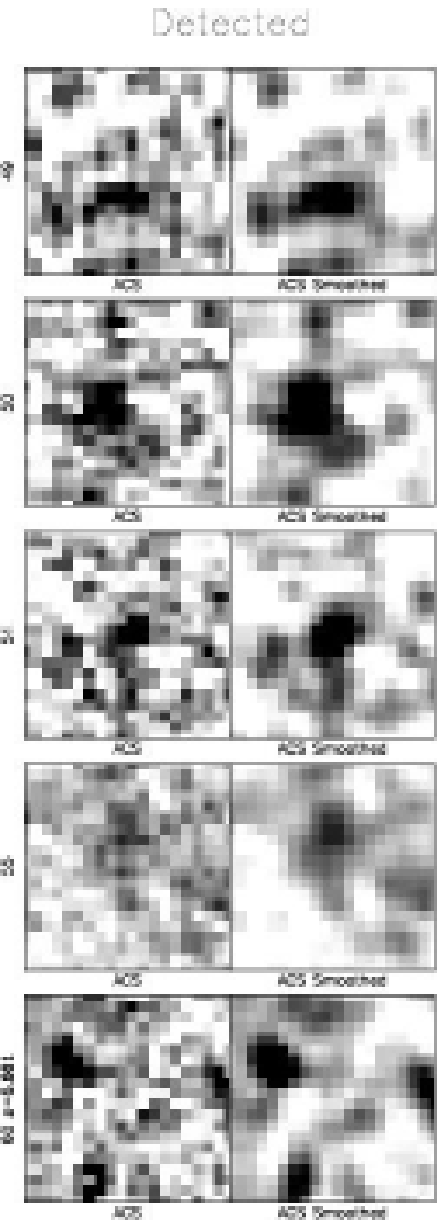}{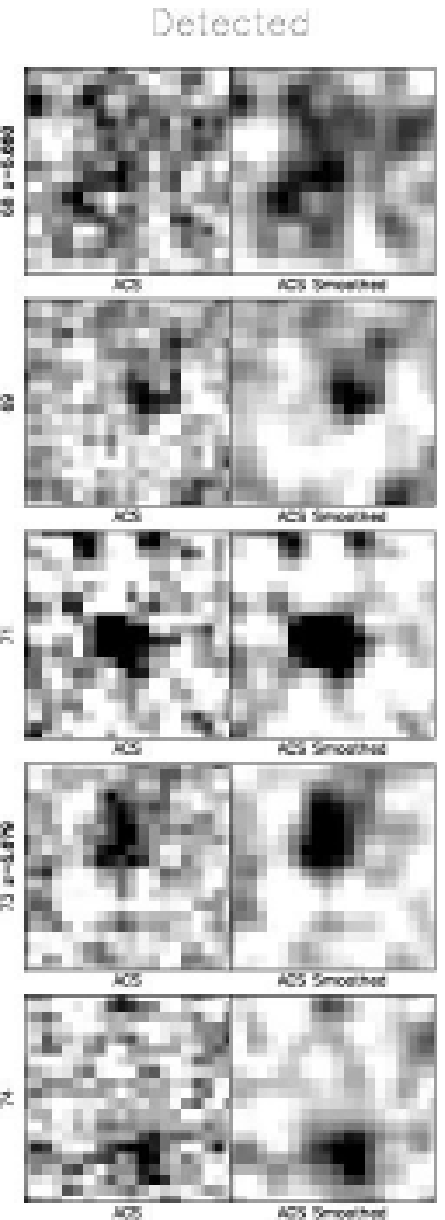}
\clearpage

\plottwo{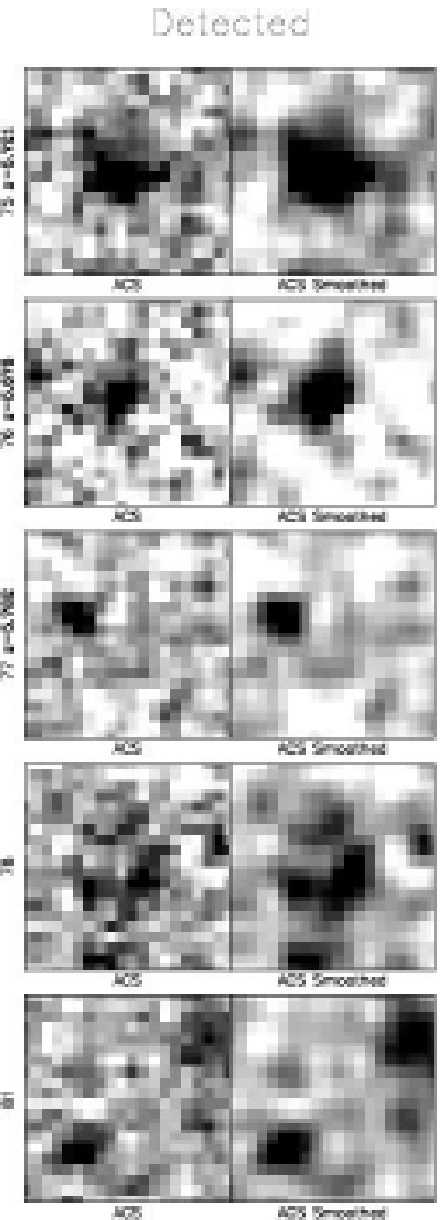}{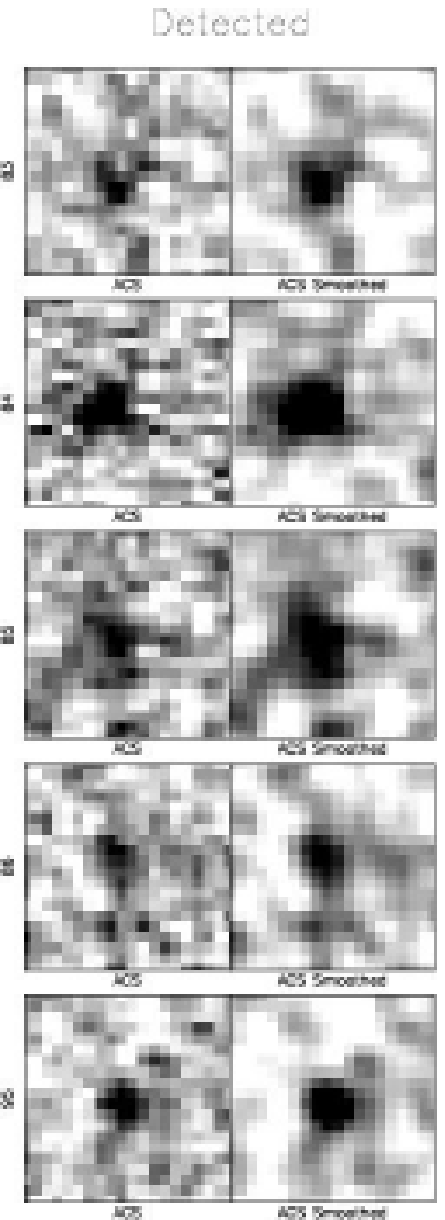}
\clearpage

\plottwo{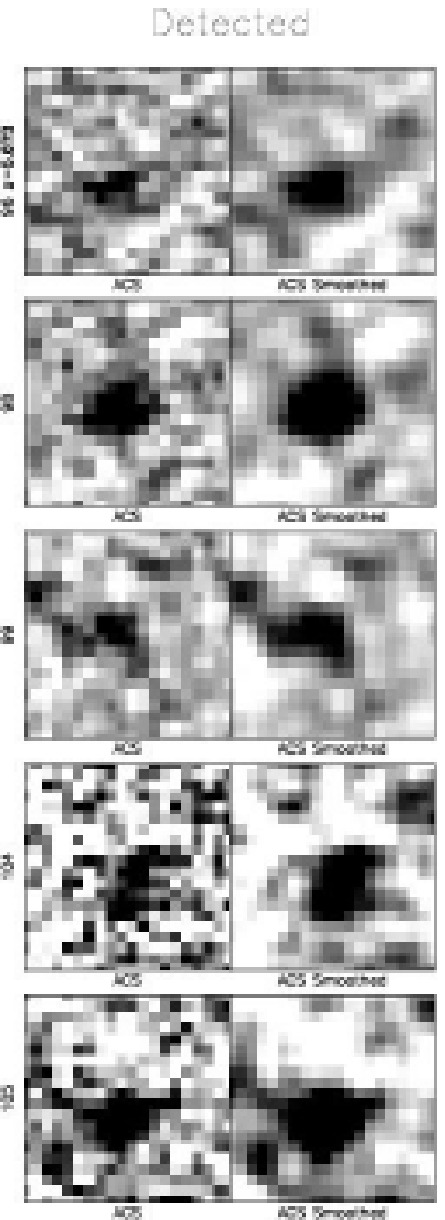}{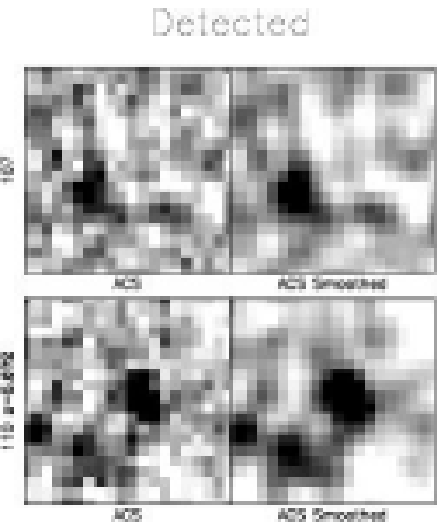}
\clearpage

\plottwo{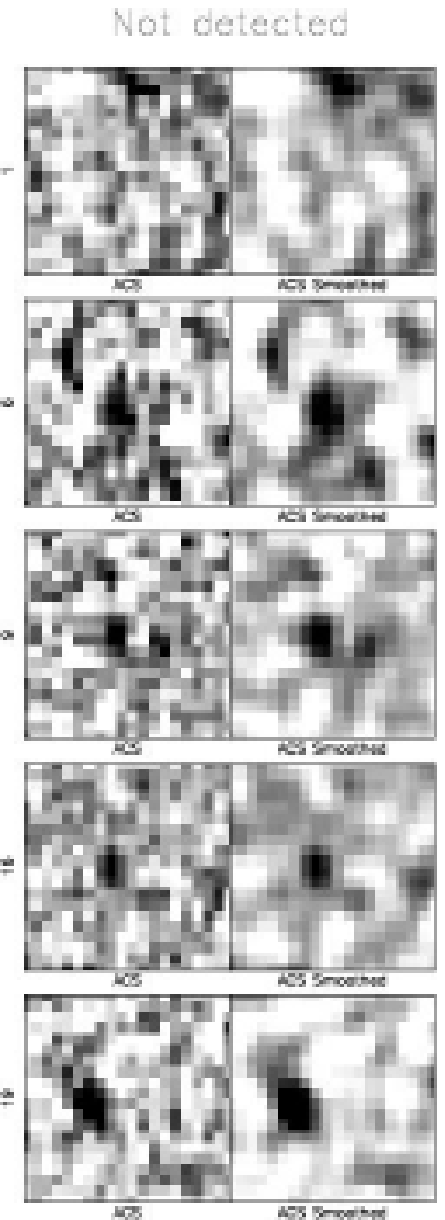}{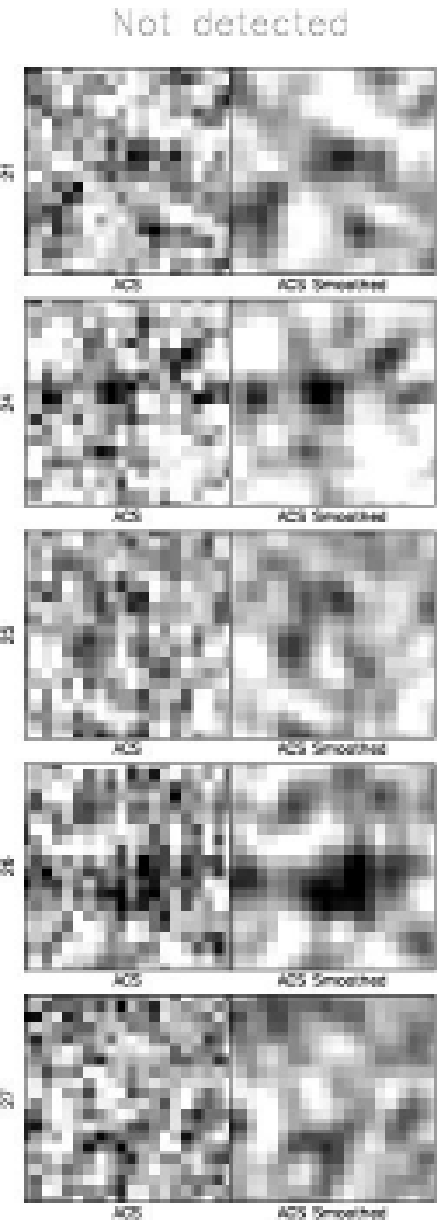}
\clearpage

\plottwo{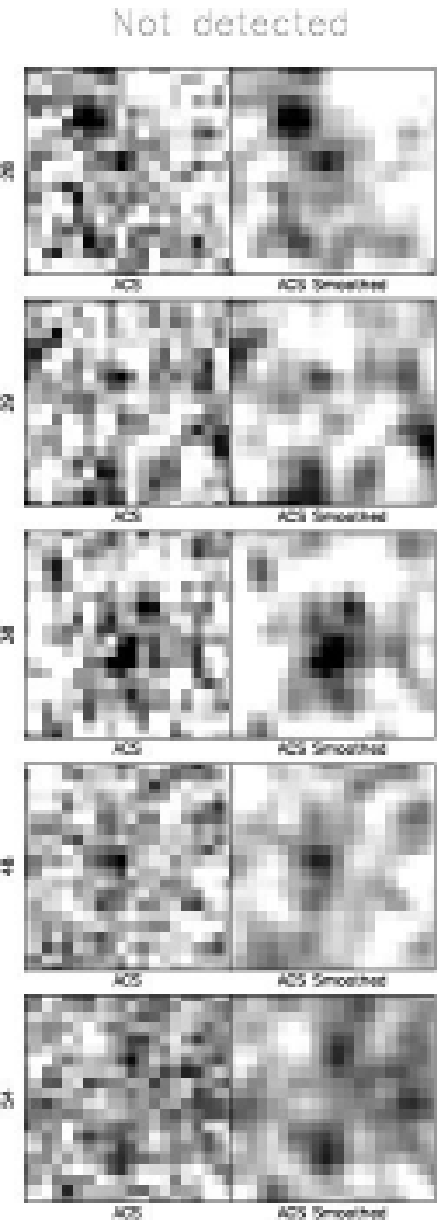}{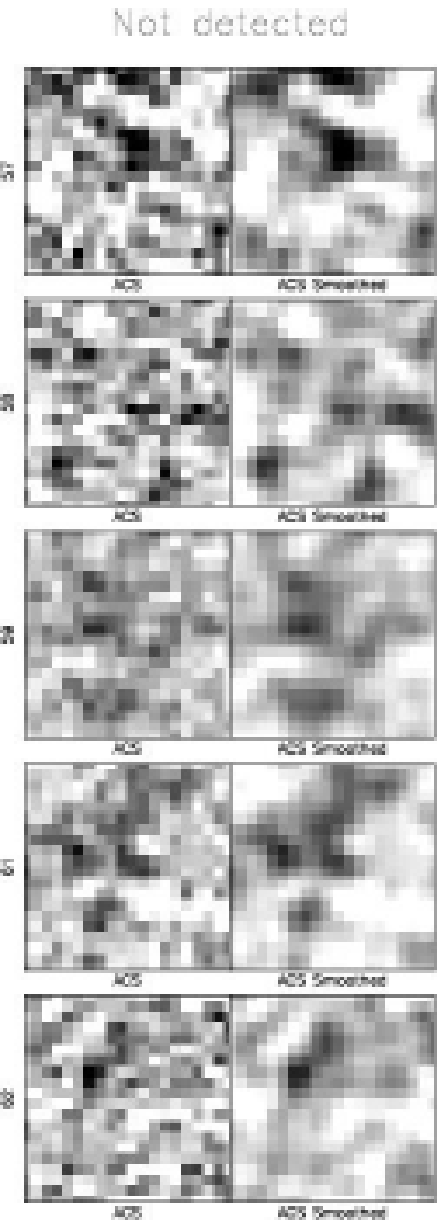}
\clearpage

\plottwo{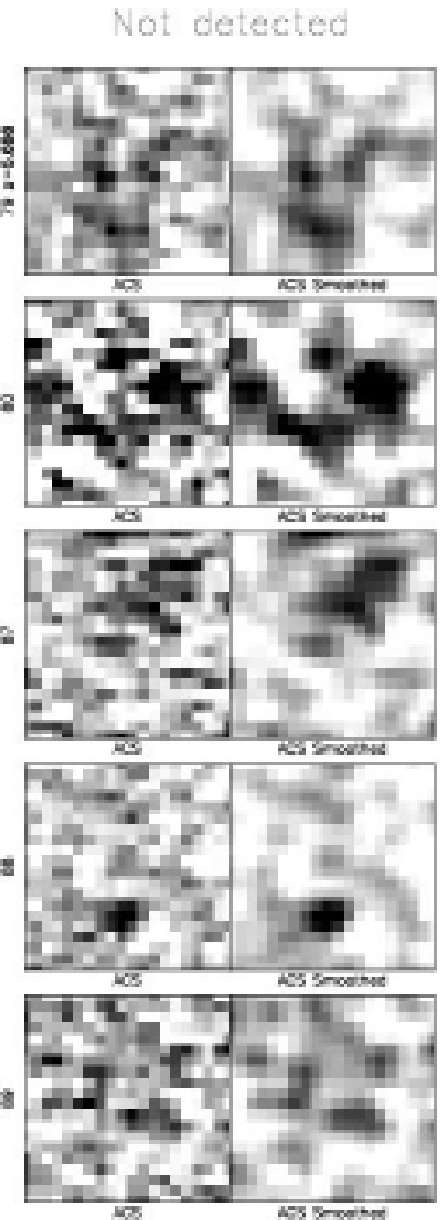}{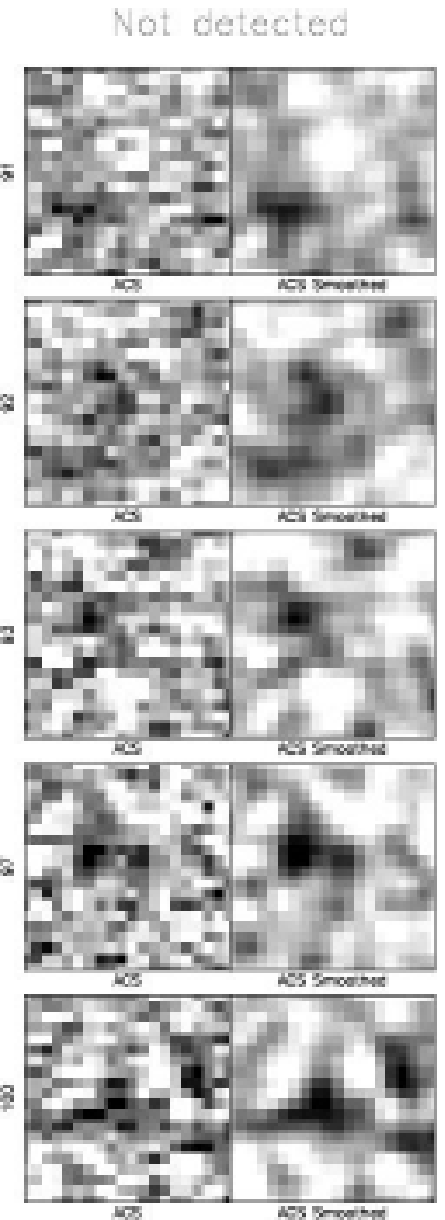}
\clearpage

\plottwo{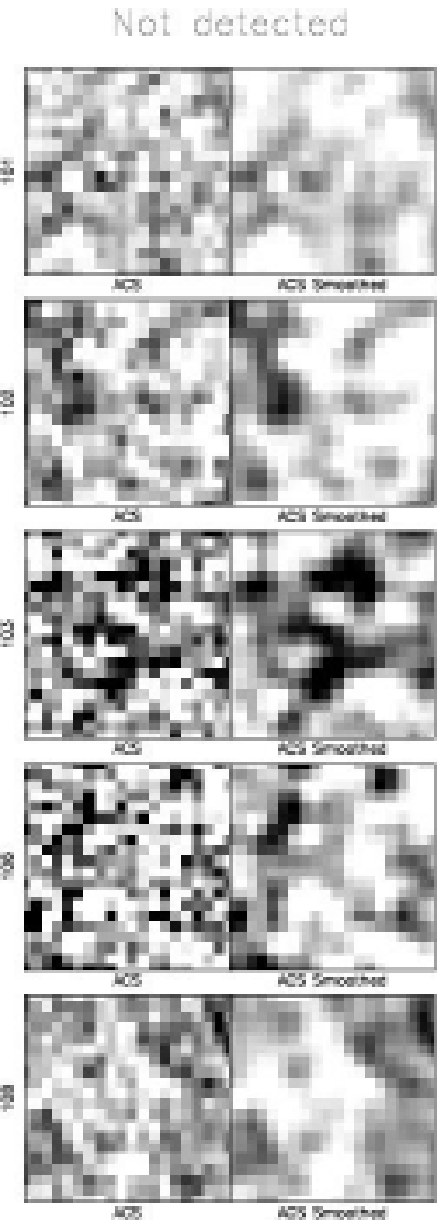}{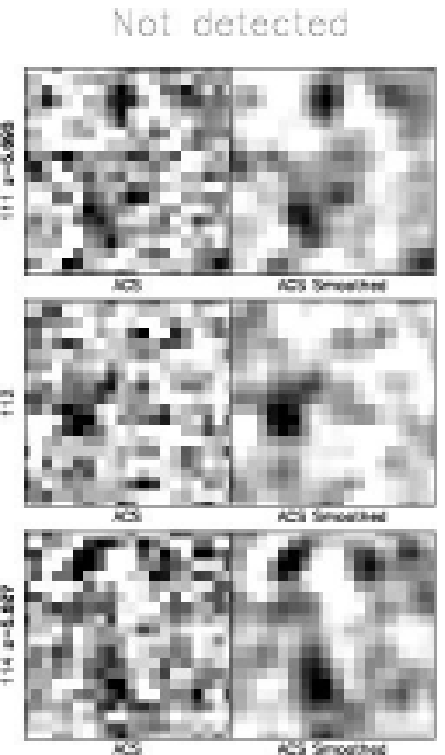}
\clearpage

\epsscale{1.0}


\begin{figure}
\epsscale{0.8}
\plotone{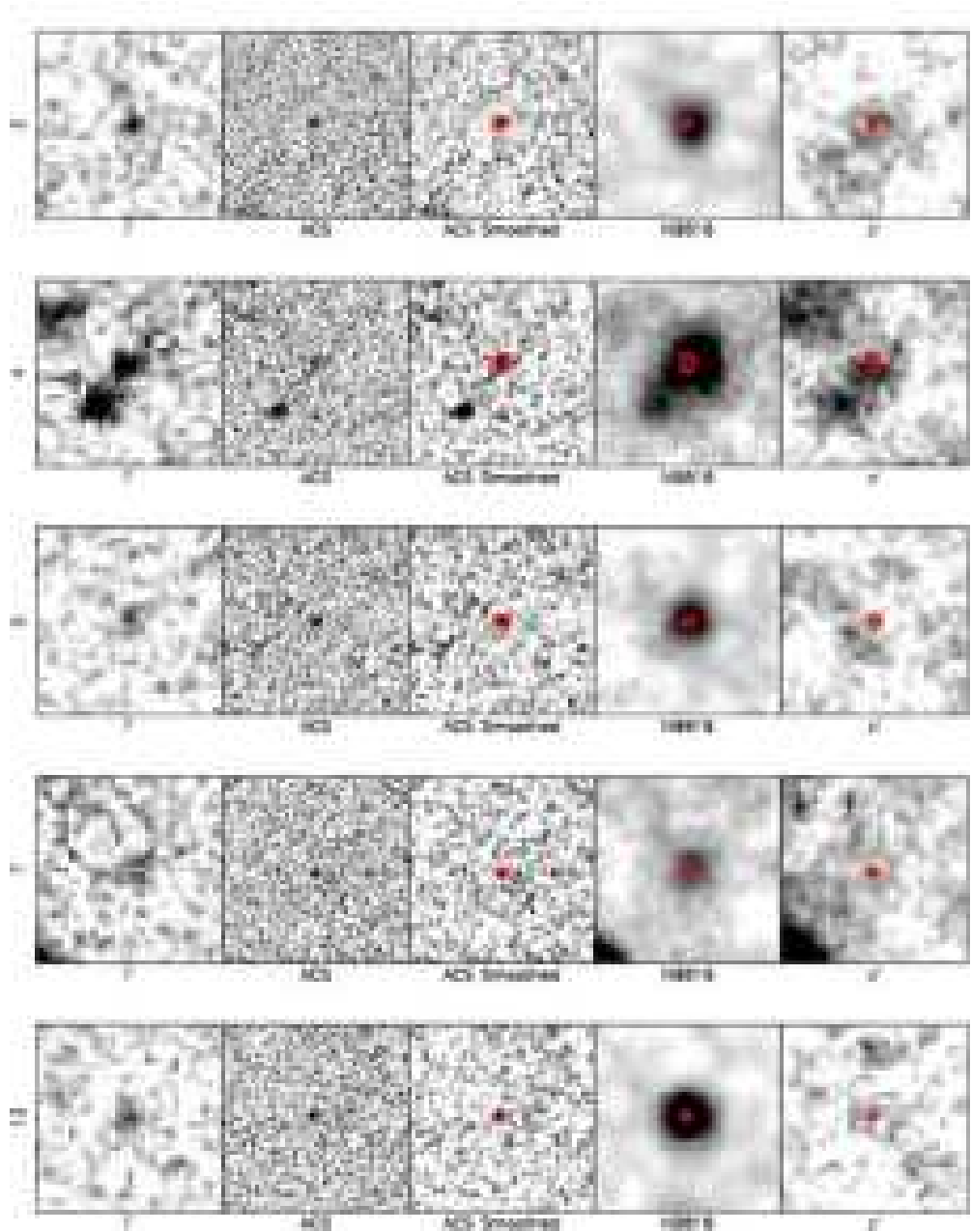}
\caption{Thumbnails of all  85 LAEs analyzed in this paper. 
Each panel has a size of 5$^{\prime\prime} \times$
5$^{\prime\prime}$. 
Red ellipses  overplotted on the  smoothed ACS, $NB816$, and $z'$ images are 
half light ellipses of the detected LAE components in the ACS image.
Green and blue ellipses are ACS sources excluded from the sample by eye inspection and
by rejection of  foreground neighbors, respectively.}
\end{figure}
\clearpage

\plotone{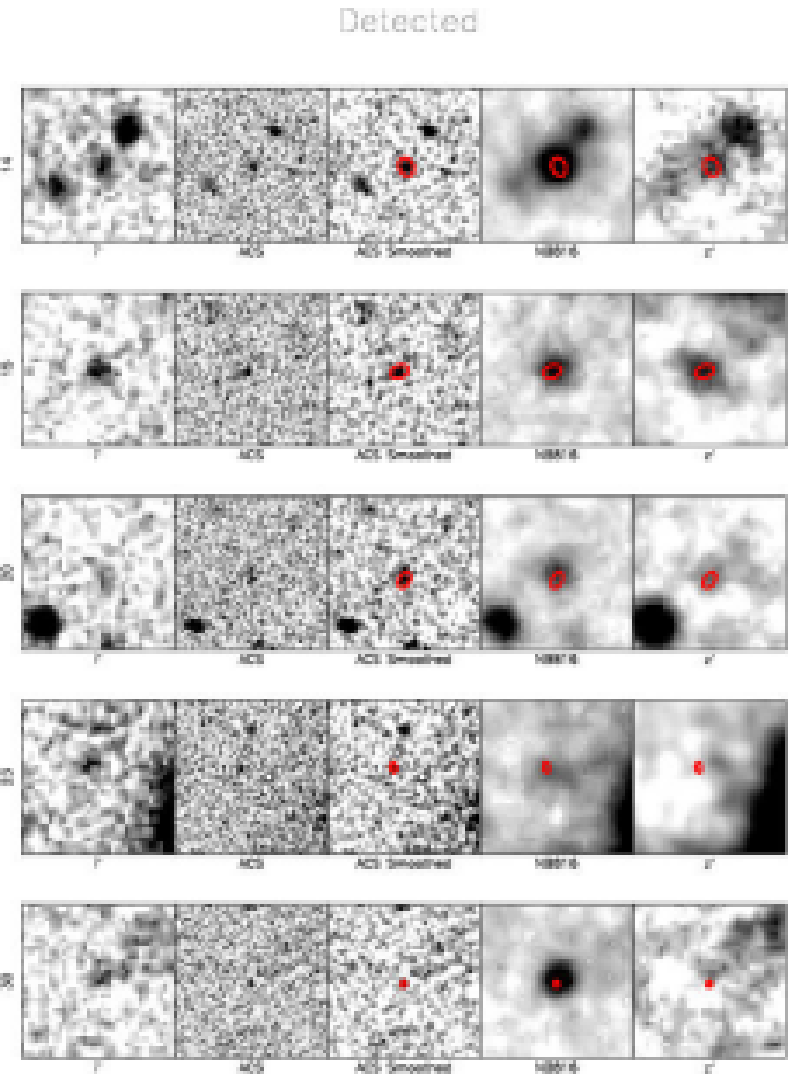}
\clearpage

\plotone{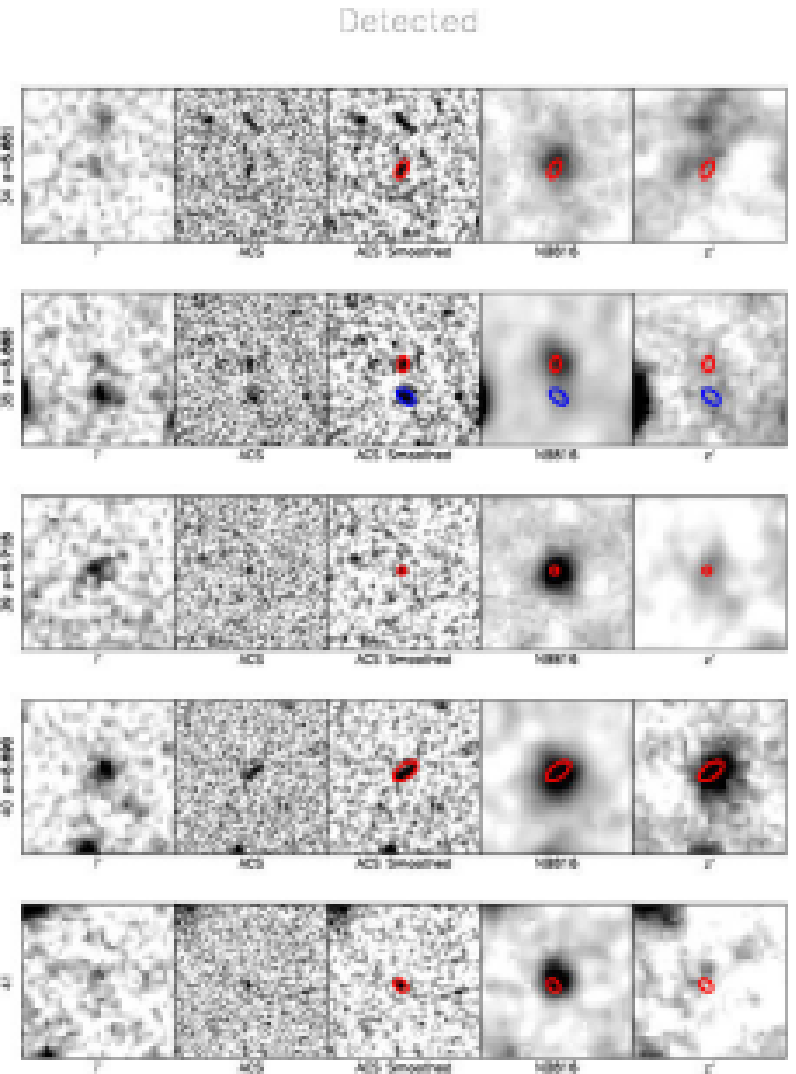}
\clearpage

\plotone{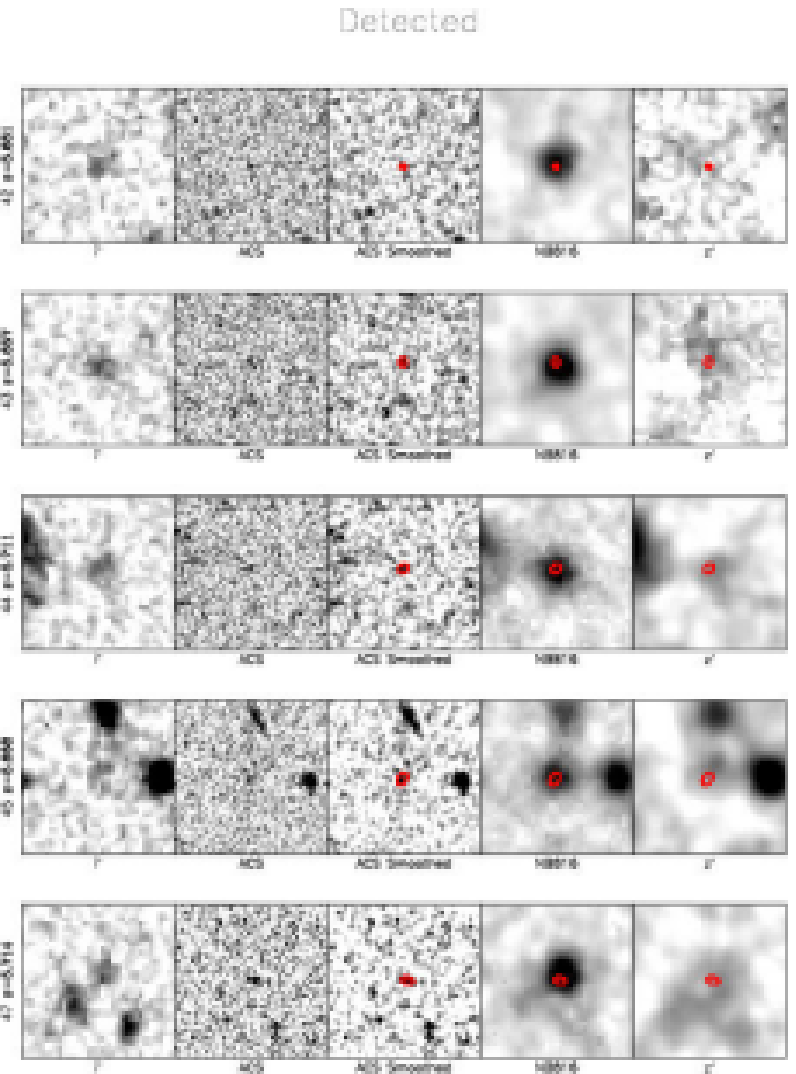}
\clearpage

\plotone{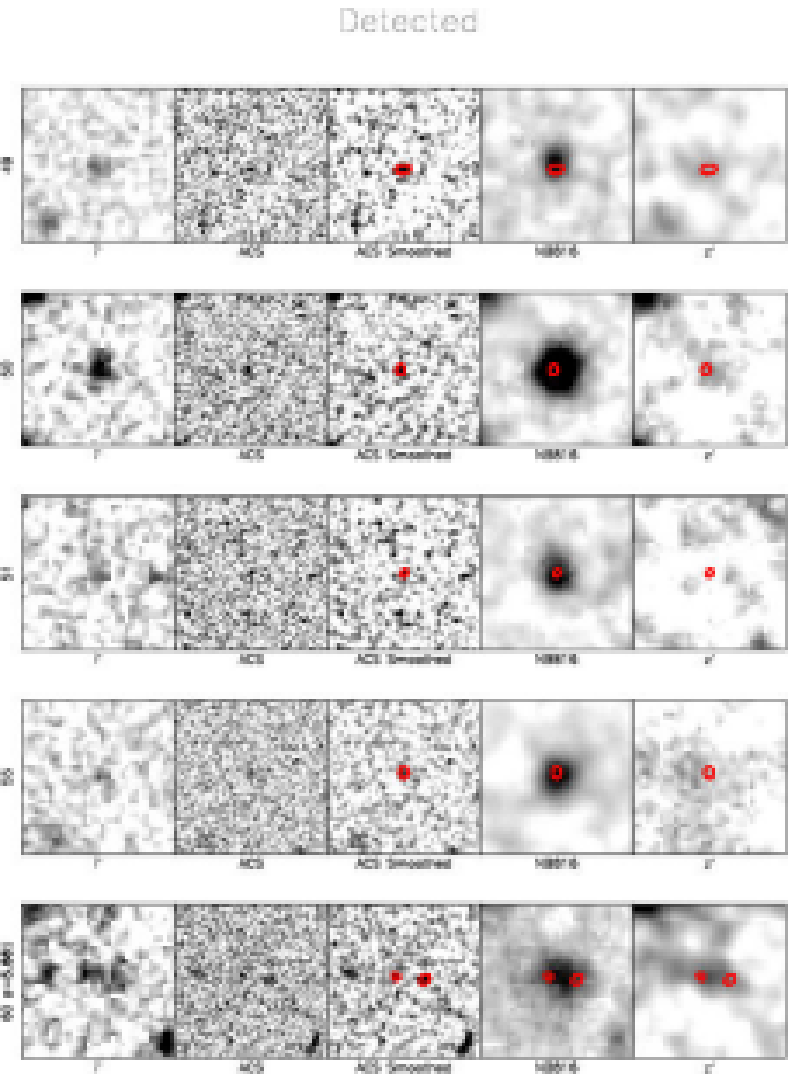}
\clearpage

\plotone{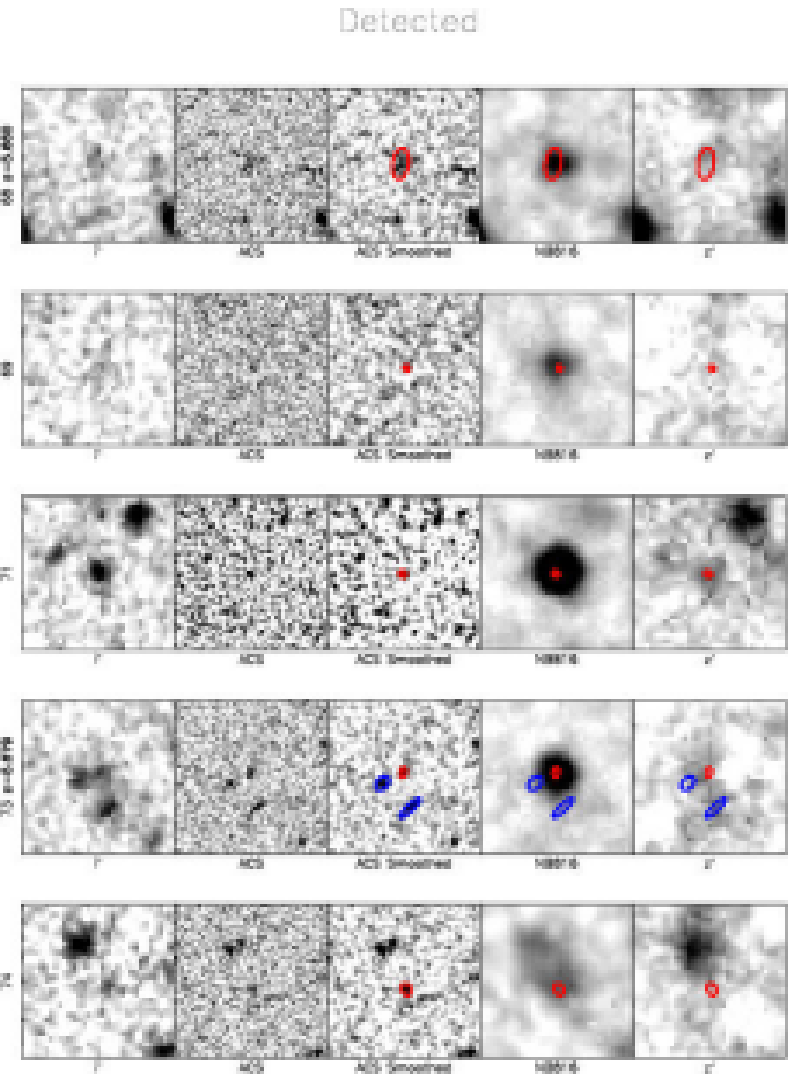}
\clearpage

\plotone{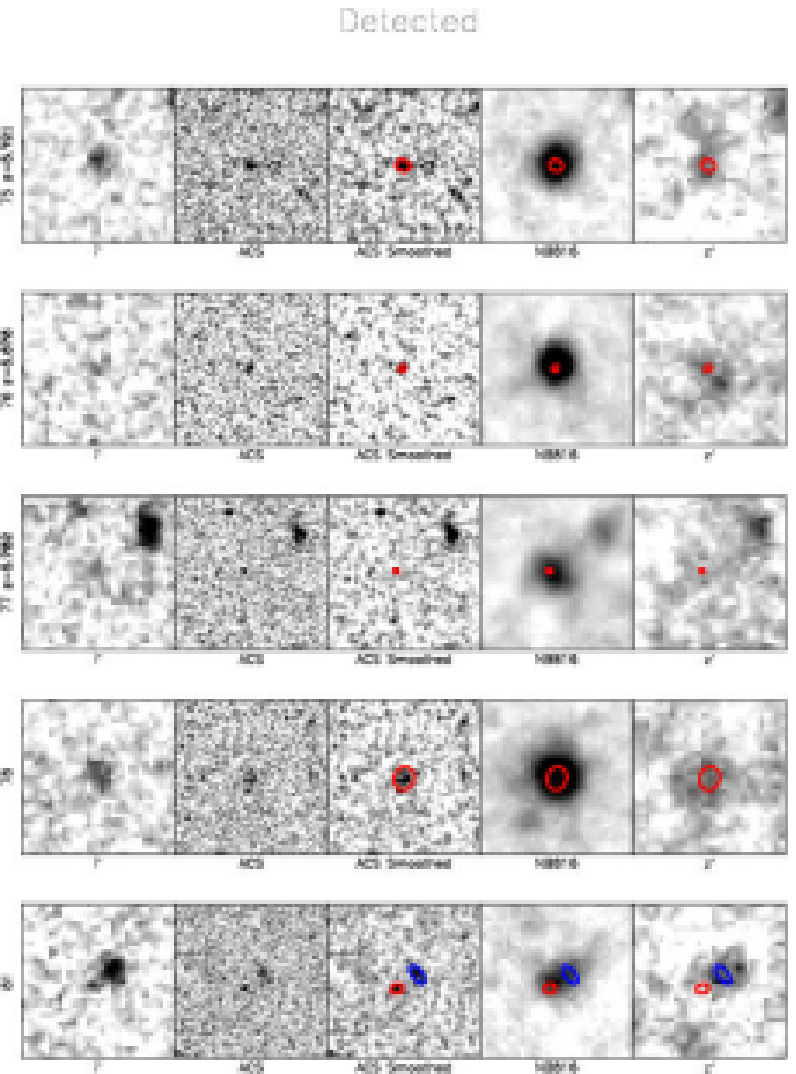}
\clearpage

\plotone{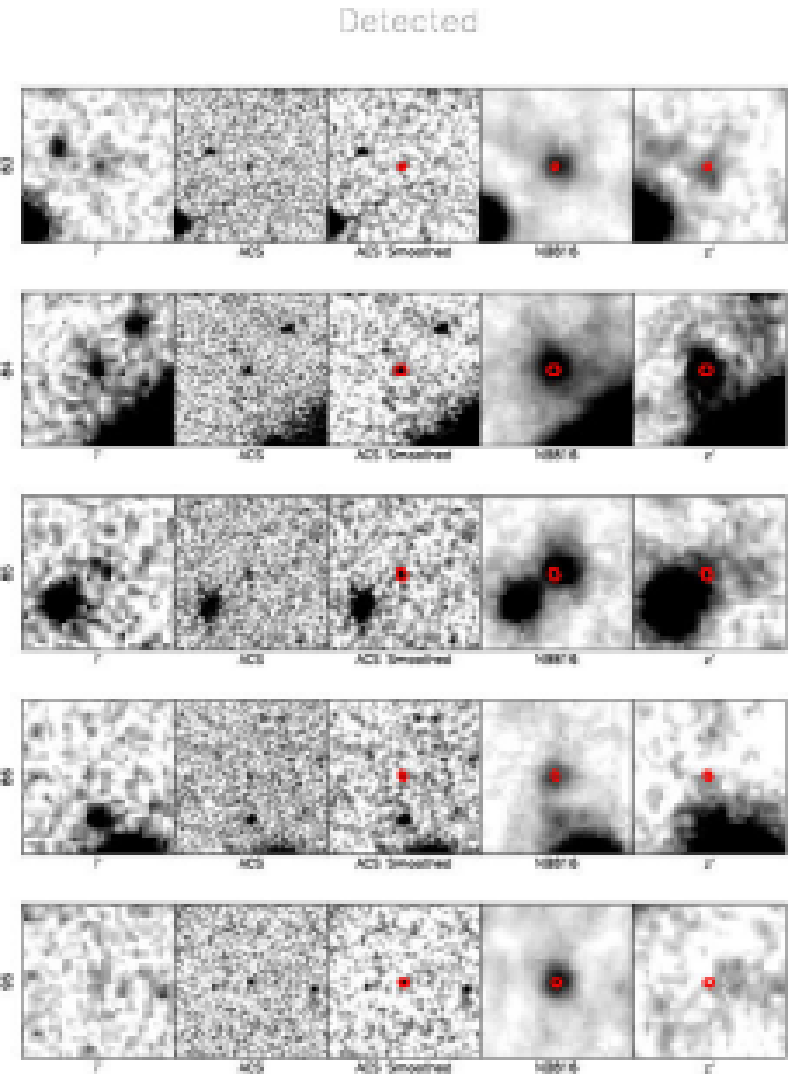}
\clearpage

\plotone{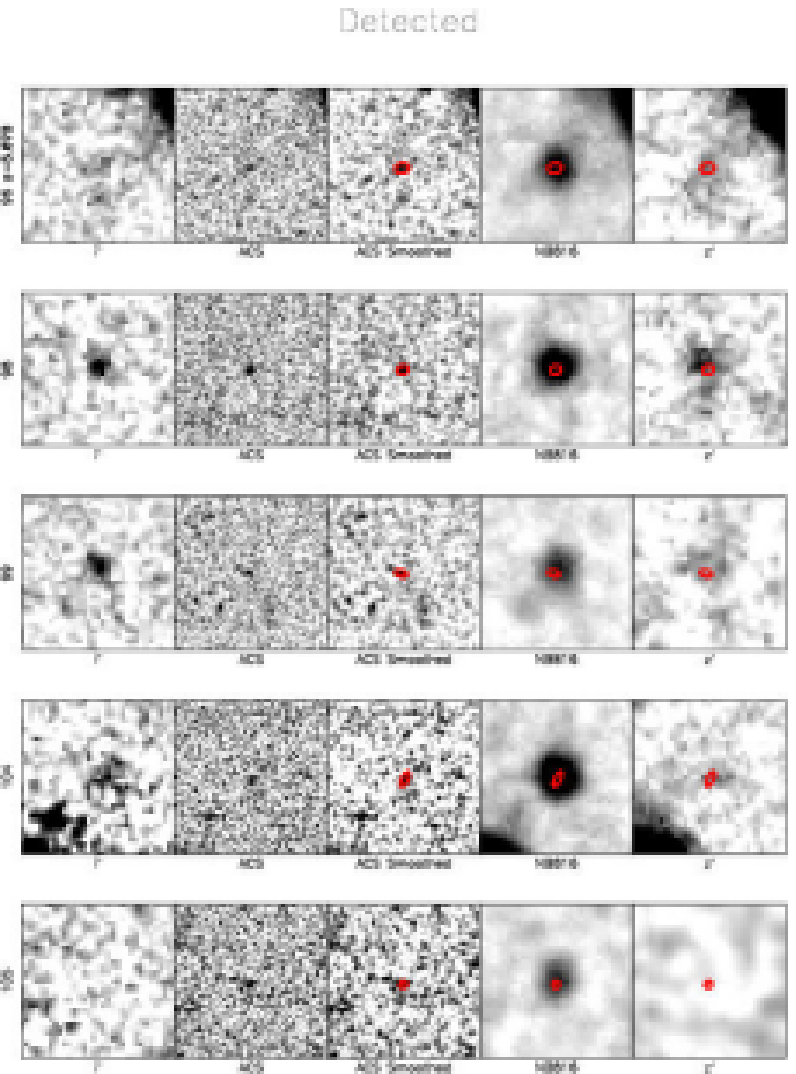}
\clearpage

\plotone{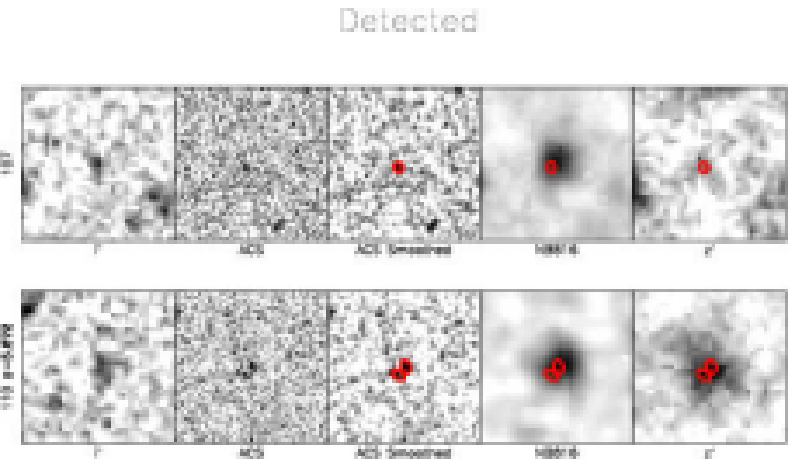}
\clearpage

\plotone{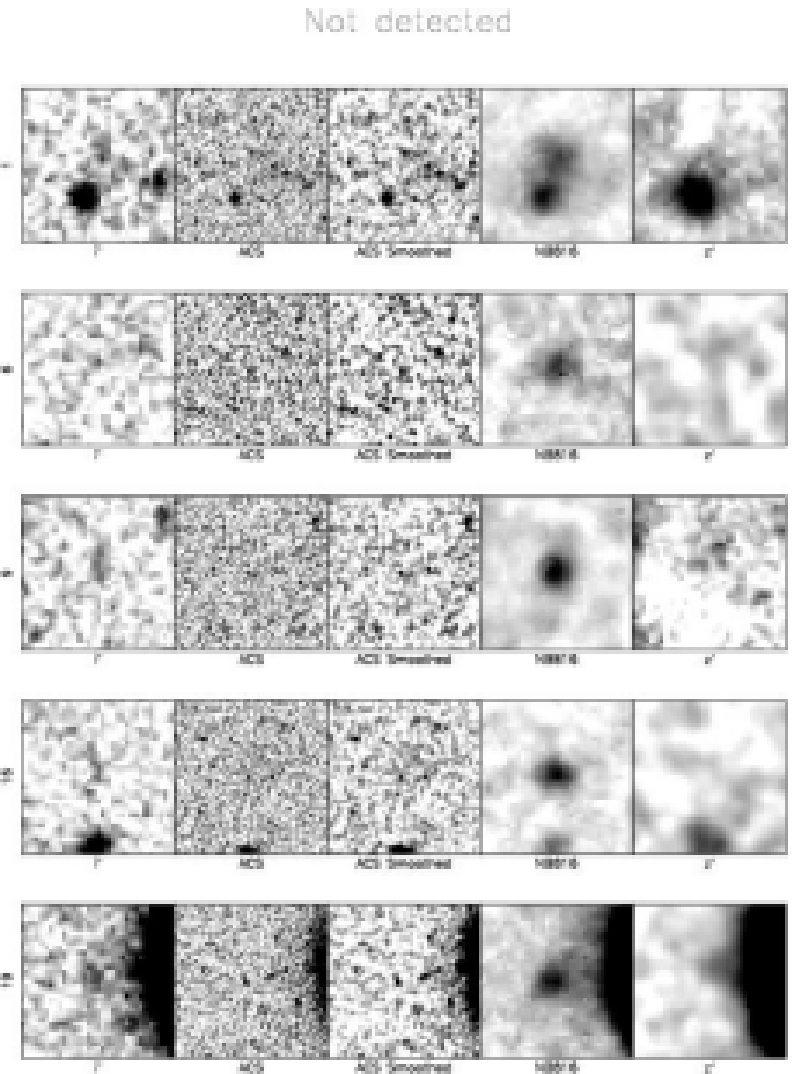}
\clearpage

\plotone{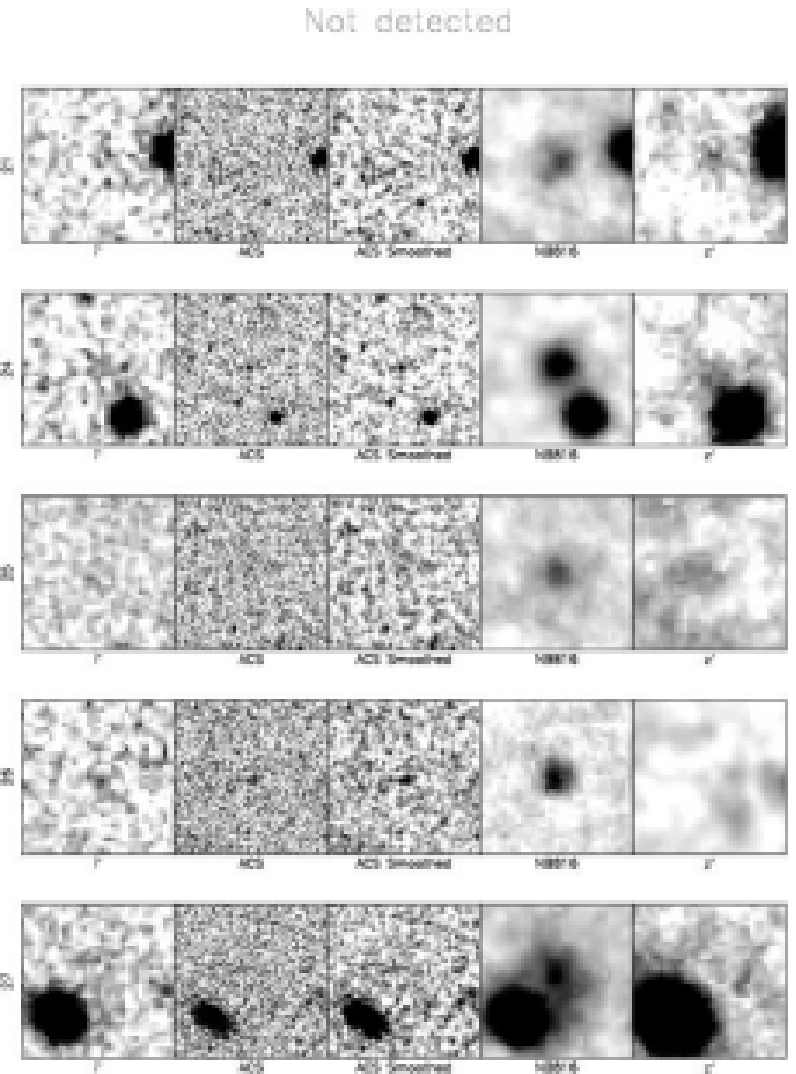}
\clearpage

\plotone{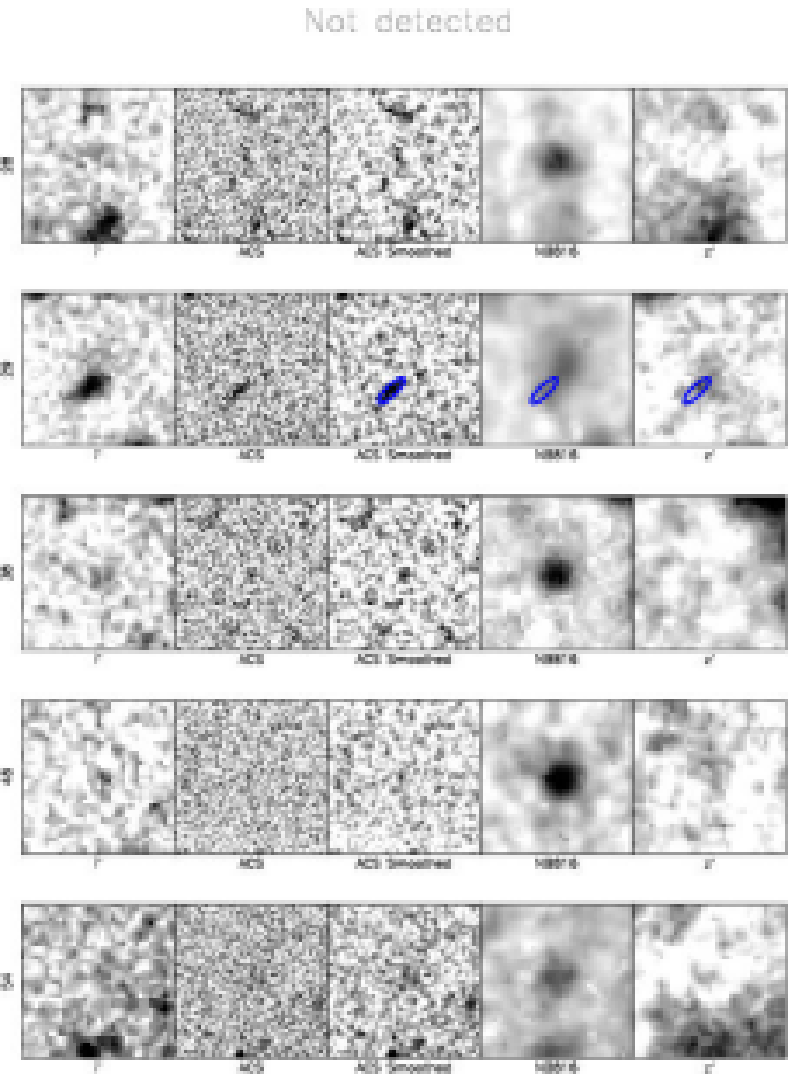}
\clearpage

\plotone{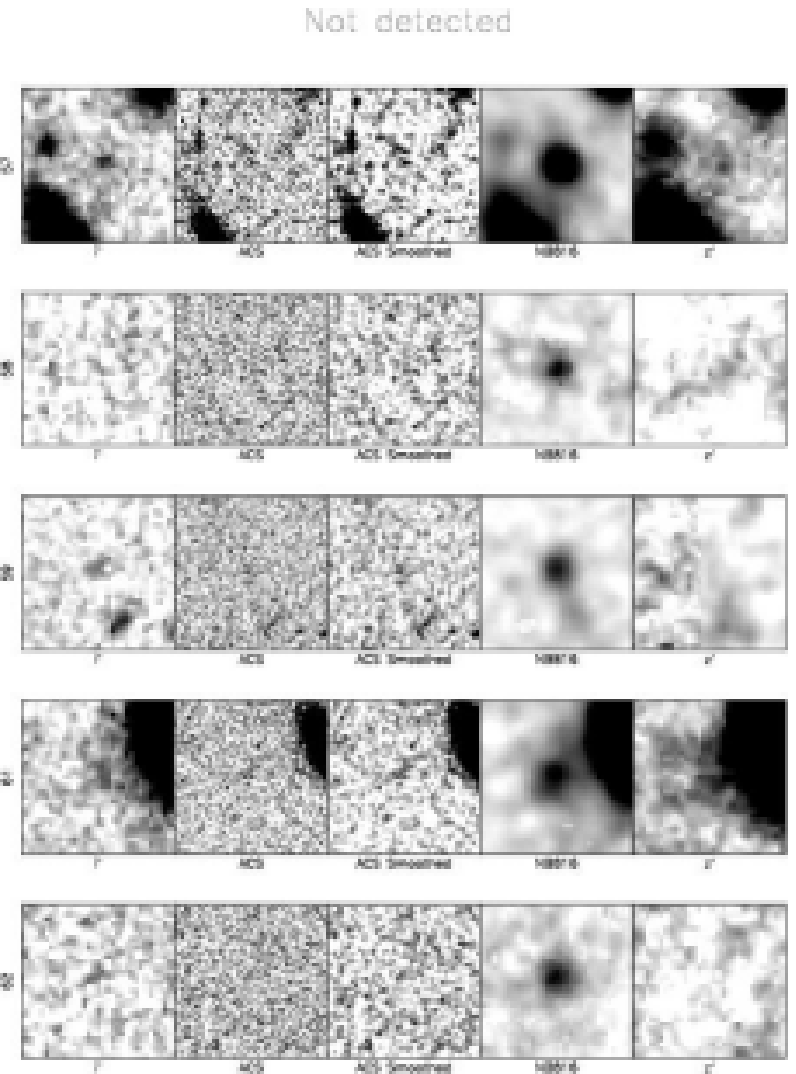}
\clearpage

\plotone{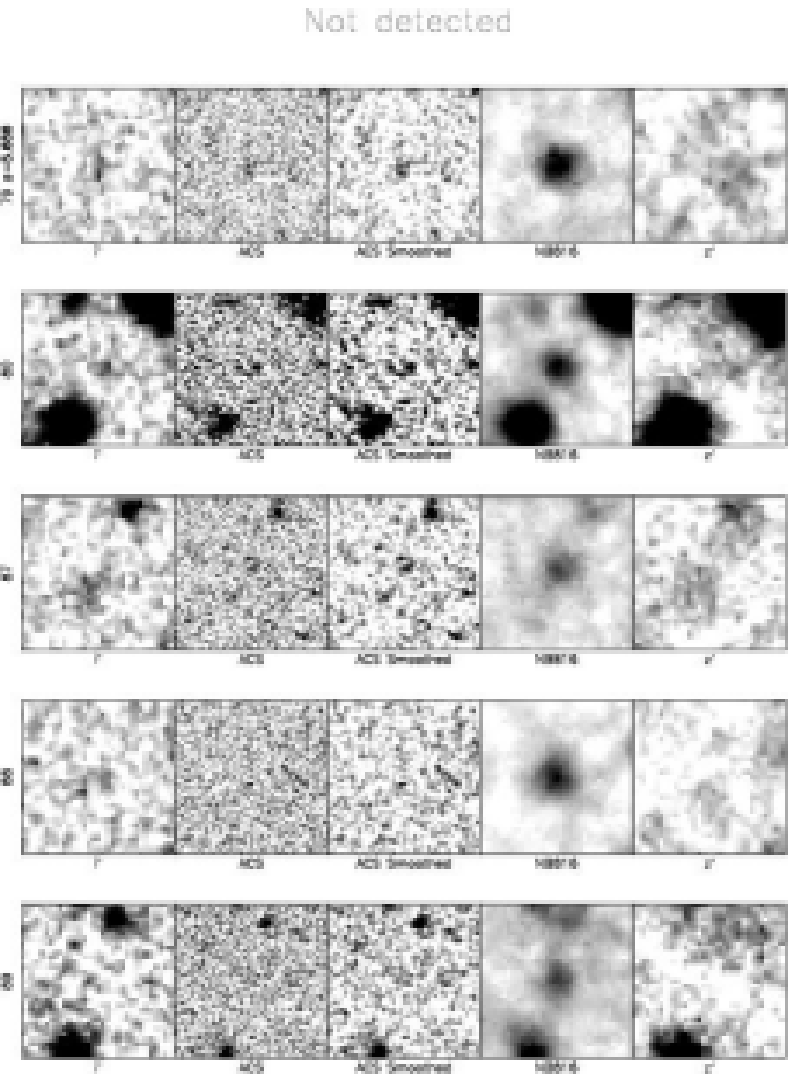}
\clearpage

\plotone{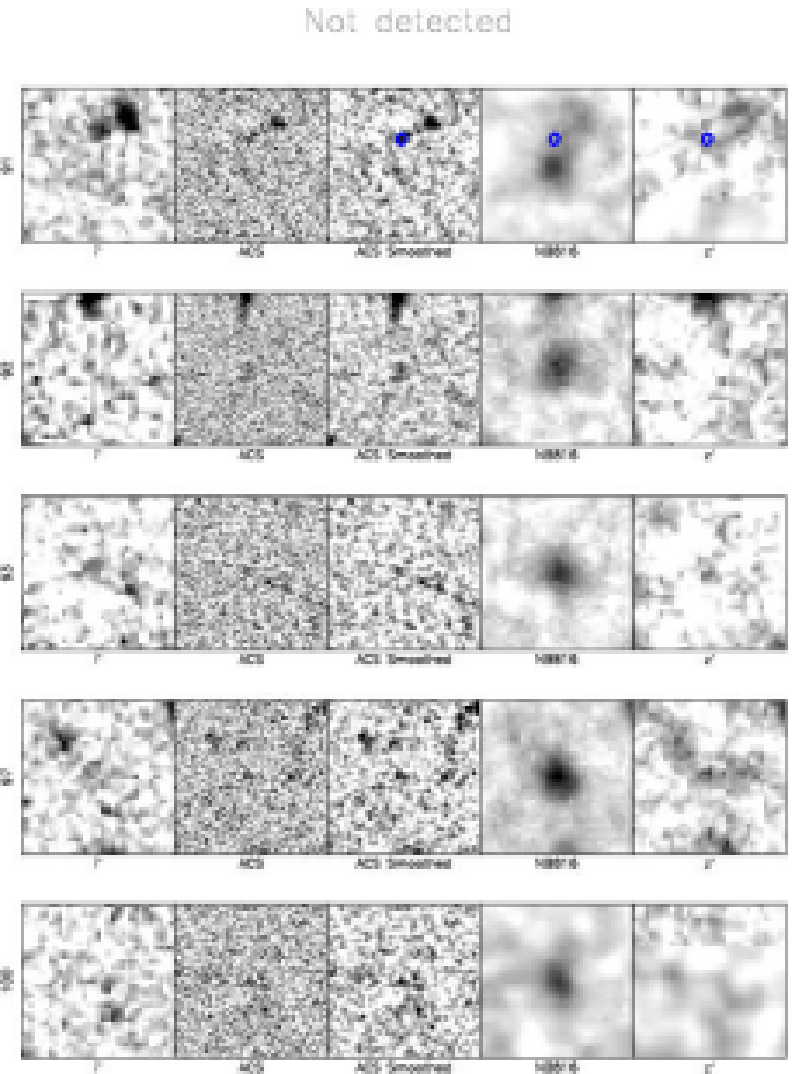}
\clearpage

\plotone{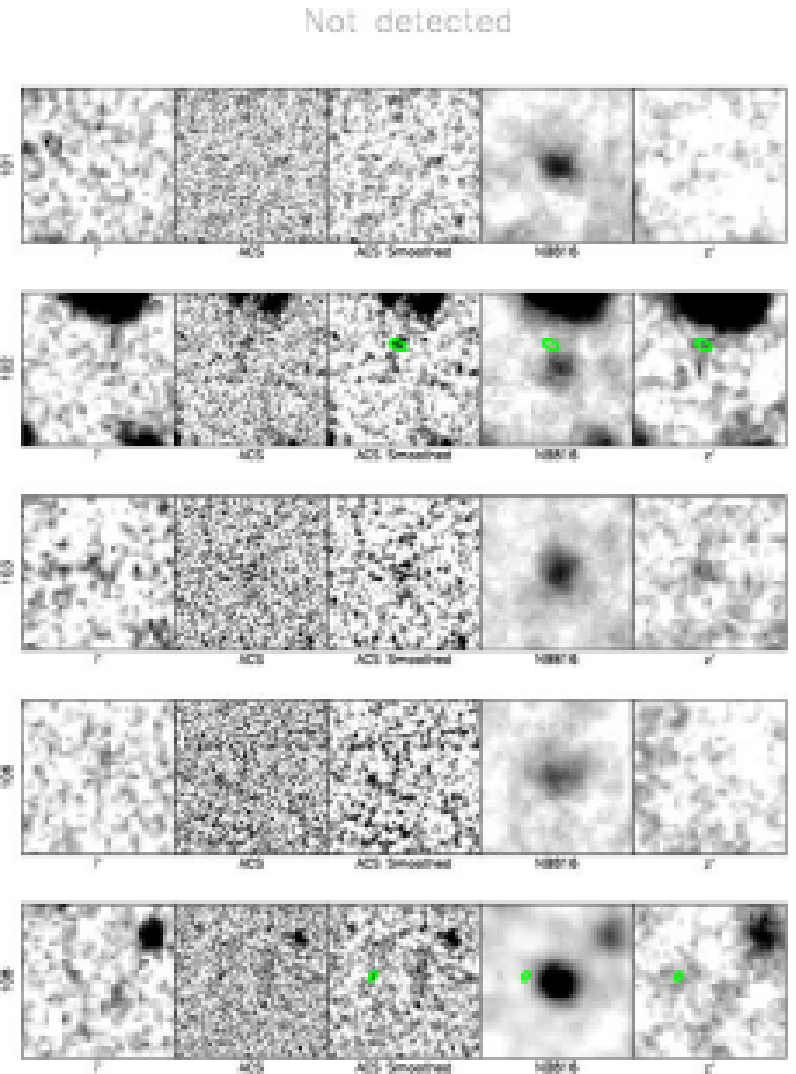}
\clearpage

\plotone{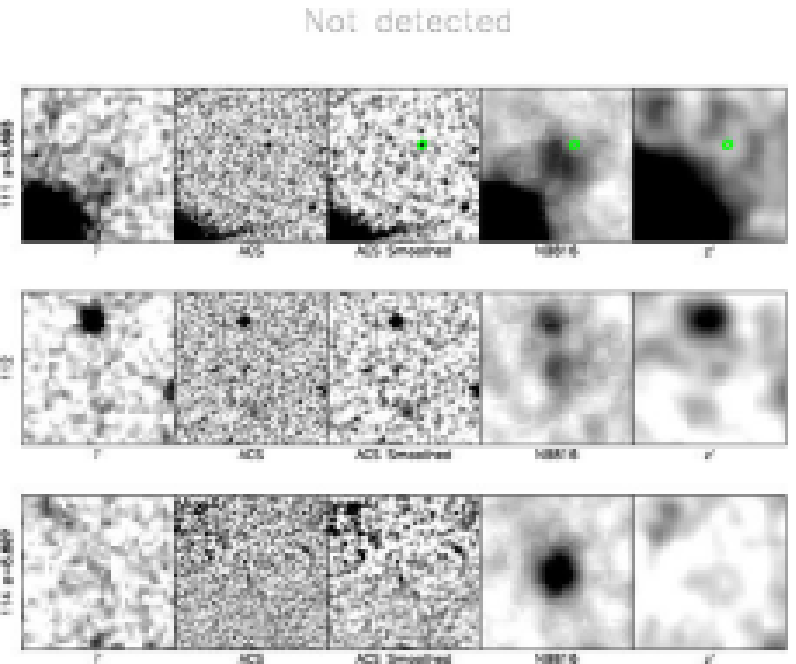}
\clearpage

\epsscale{1.0}


\begin{figure}
\plotone{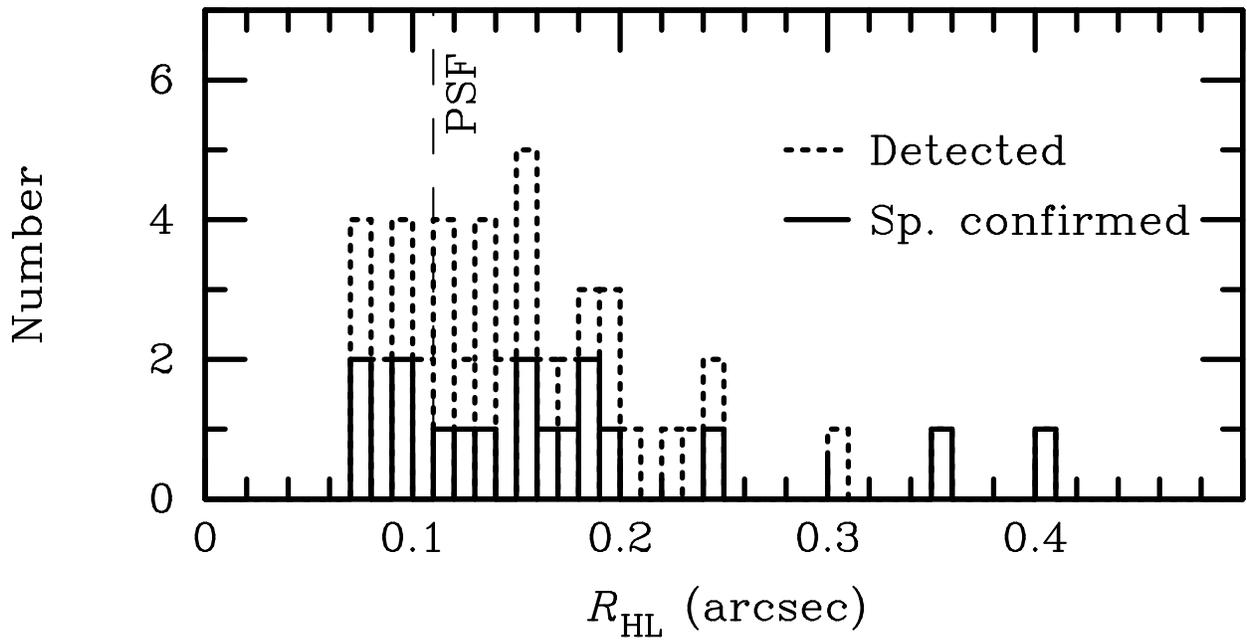}
\caption{Distribution of the half-light radius in the ACS images, $R_{\rm HL}$,
for the 47 LAEs detected with ACS by a dotted histogram.
LAEs with spectroscopic confirmation are shown by a solid histogram.}
The PSF size (0.11 arcsec) derived from stars
is shown by the dashed line. 
\end{figure}
\clearpage


\begin{figure}
\plotone{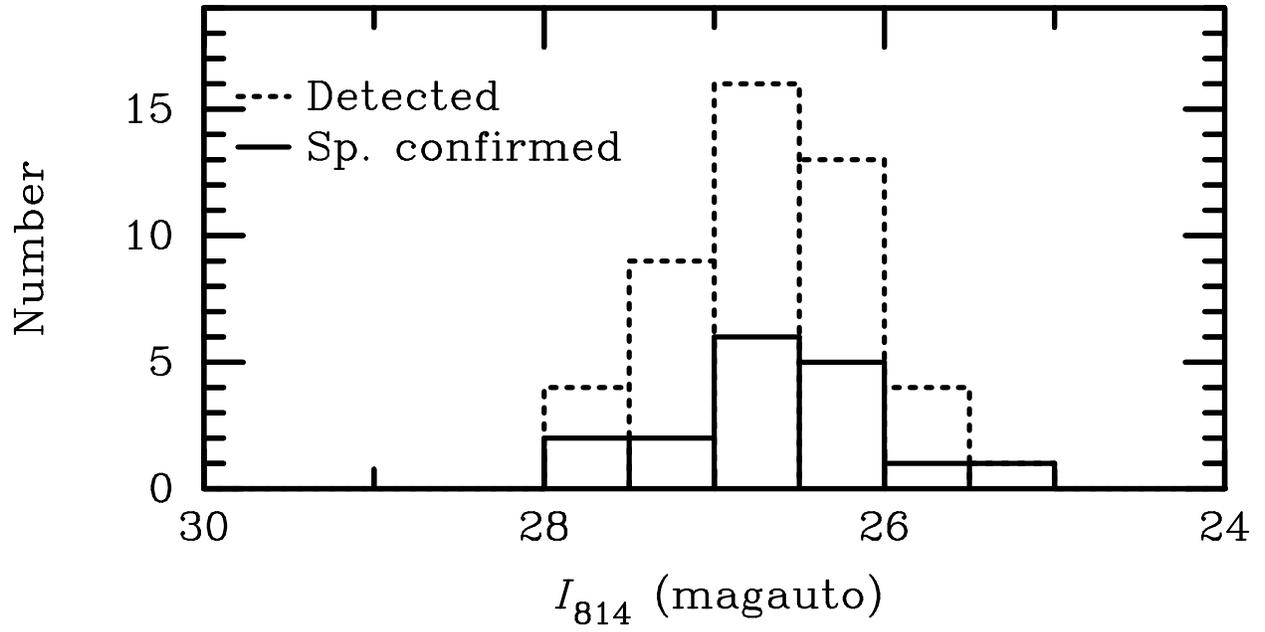}
\caption{Frequency distribution of the ACS magnitude $I_{814}$
by a dotted histogram. 
LAEs confirmed by our follow-up
spectroscopy are shown as a solid histogram. }
\end{figure}
\clearpage


\begin{figure}
\plotone{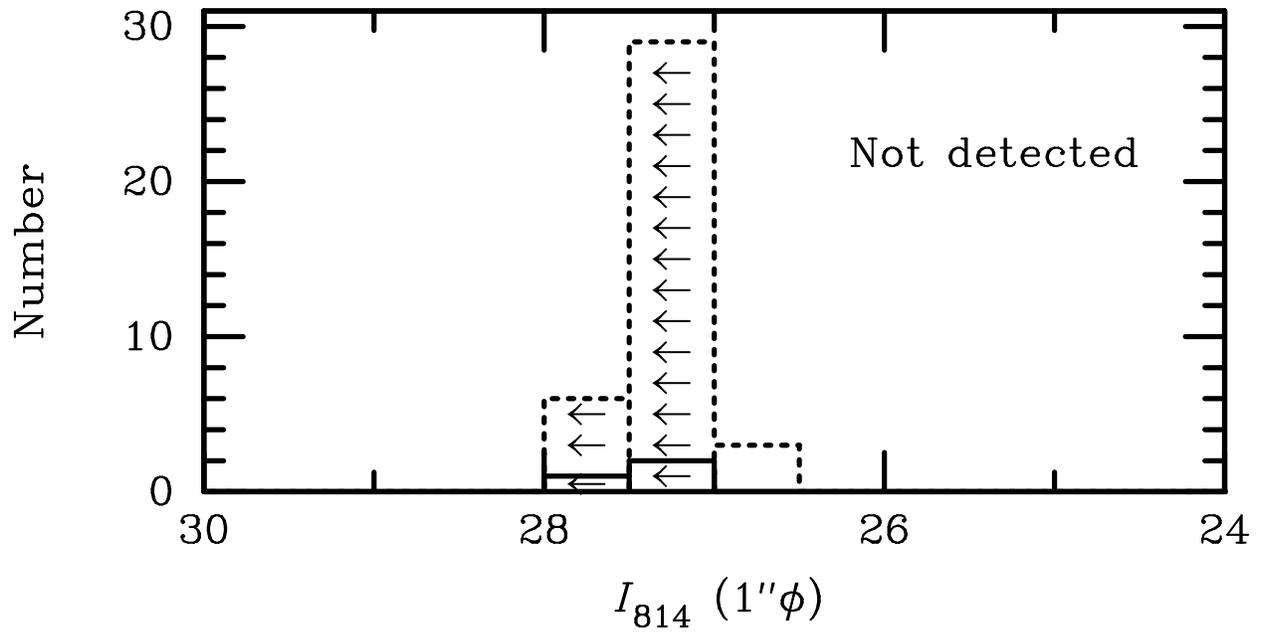}
\caption{Frequency distribution of the ACS magnitude $I_{814}$ for the not-detected LAEs (dotted histogram);
3$\sigma$ upper limits are given for each LAE. LAEs confirmed by our follow-up
spectroscopy are shown by a solid histogram. }
\end{figure}
\clearpage


\begin{figure}
\plotone{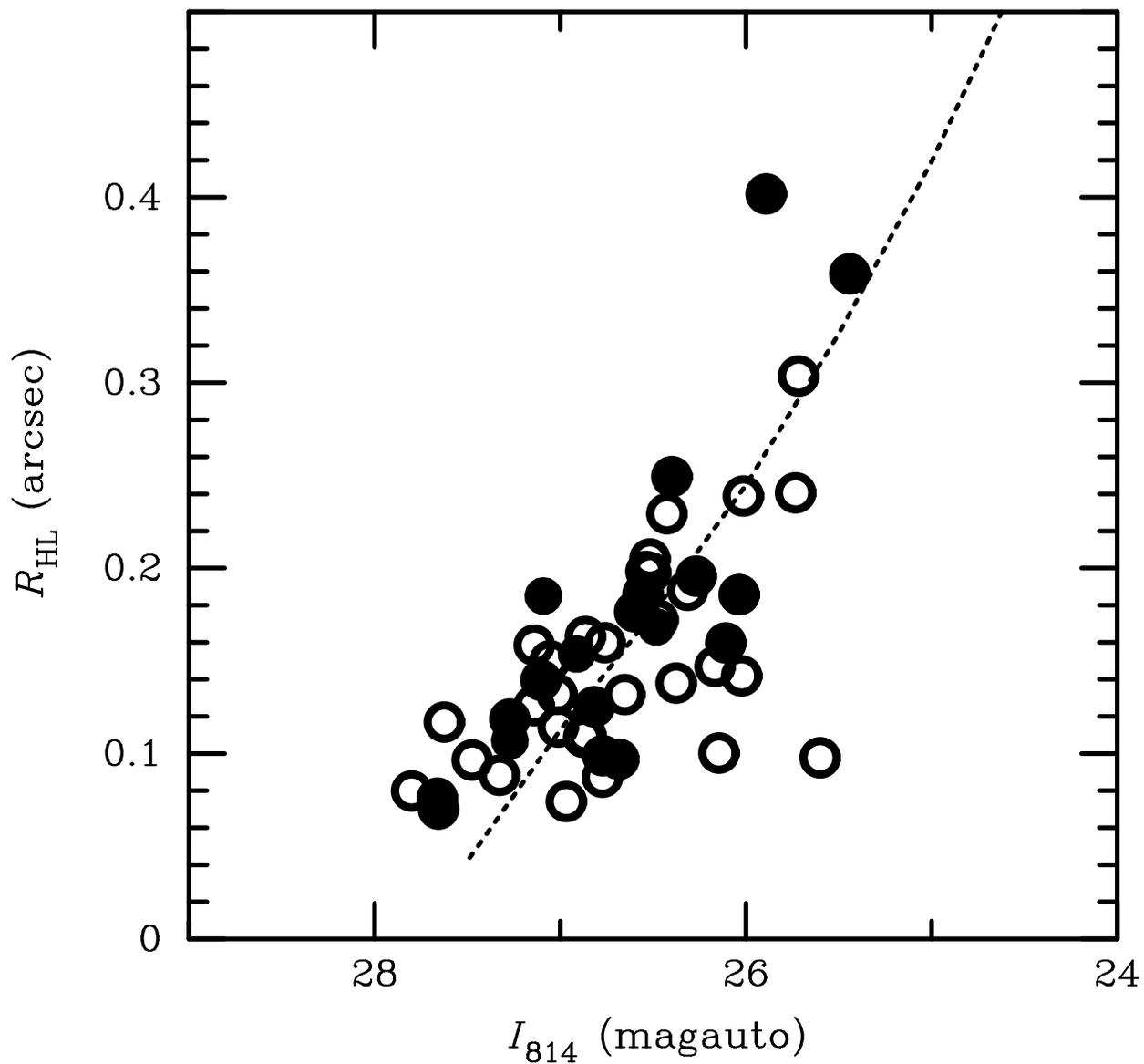}
\caption{Diagram between half-light radius $R_{\rm HL}$ and the ACS magnitude $I_{814}$.
The LAEs confirmed by our follow-up spectroscopy are shown by filled circle.
For double-component LAEs,  each component is plotted.
The 50\% detection completeness for exponential disk objects
estimated by Monte Carlo simulation is indicated by the dashed curve.
}
\end{figure}
\clearpage


\begin{figure}
\plotone{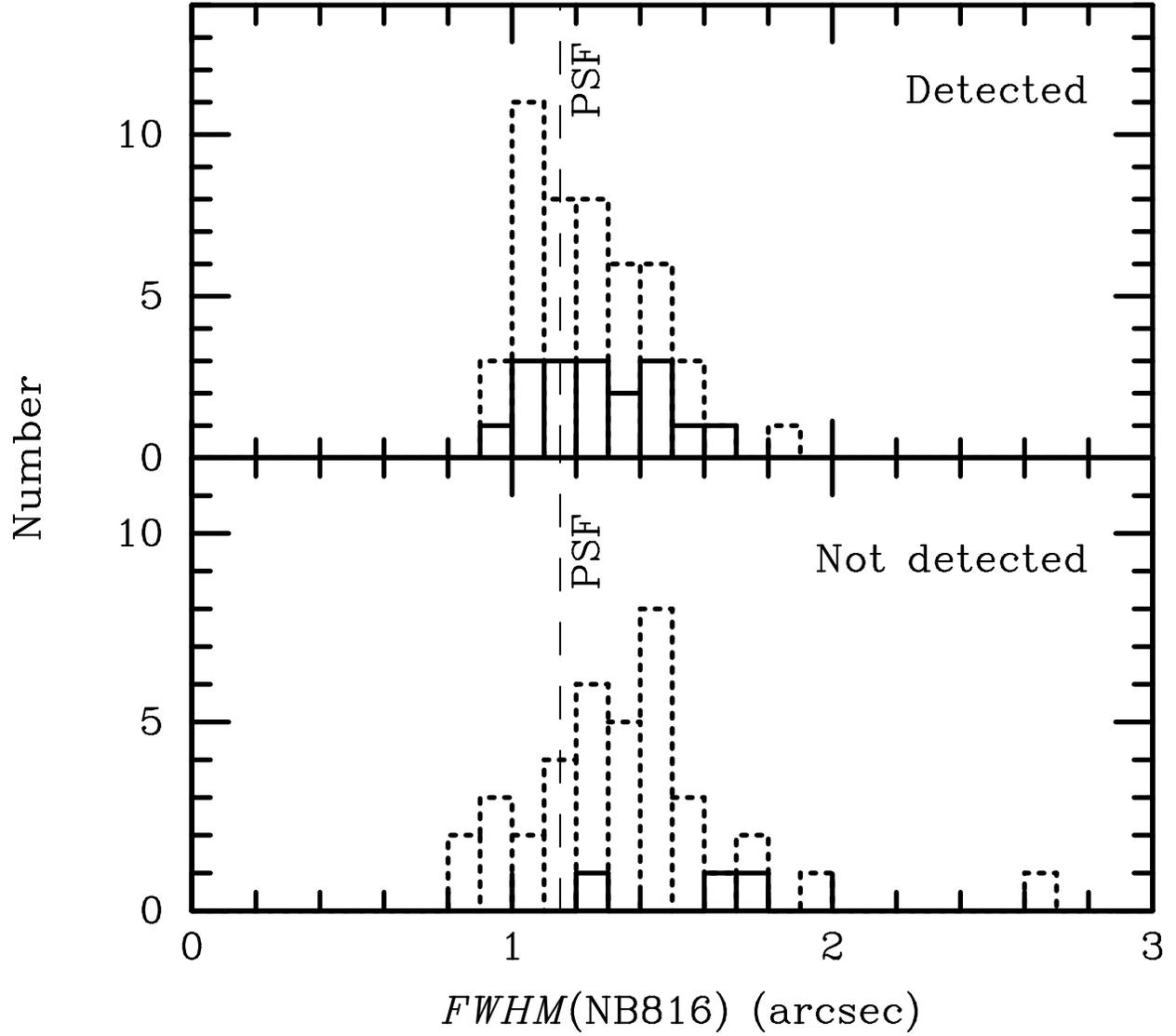}
\caption{Frequency distributions of size in the narrow-band filter $NB816$
$FWHM$(NB816) for the detected and not-detected LAEs by dotted histograms.
LAEs confirmed by our follow-up
spectroscopy are shown by solid histograms.}
\end{figure}
\clearpage


\begin{figure}
\plotone{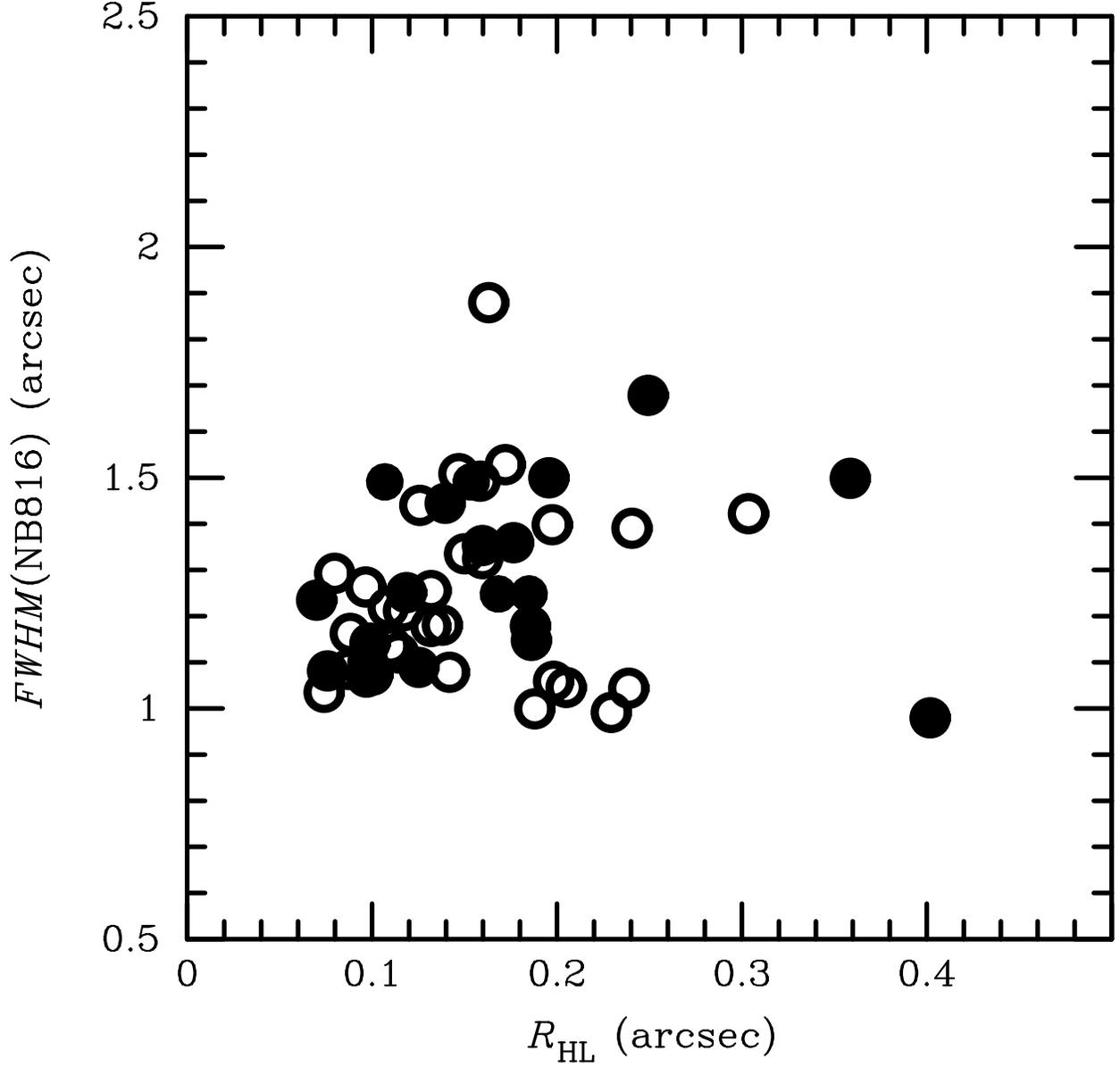}
\caption{Diagram between  size in the narrow-band filter $NB816$, $FWHM$(NB816) and the half-light radius in the ACS image,
$R_{\rm HL}$.
LAEs confirmed by our follow-up spectroscopy are shown by filled circles.
For double-component LAEs $R_{\rm HL}$ is plotted for each component
while the same  $FWHM$(NB816) is adopted for the two components
as they are not resolved in the NB816 images.
}
\end{figure}
\clearpage


\begin{figure}
\epsscale{0.5}
\plotone{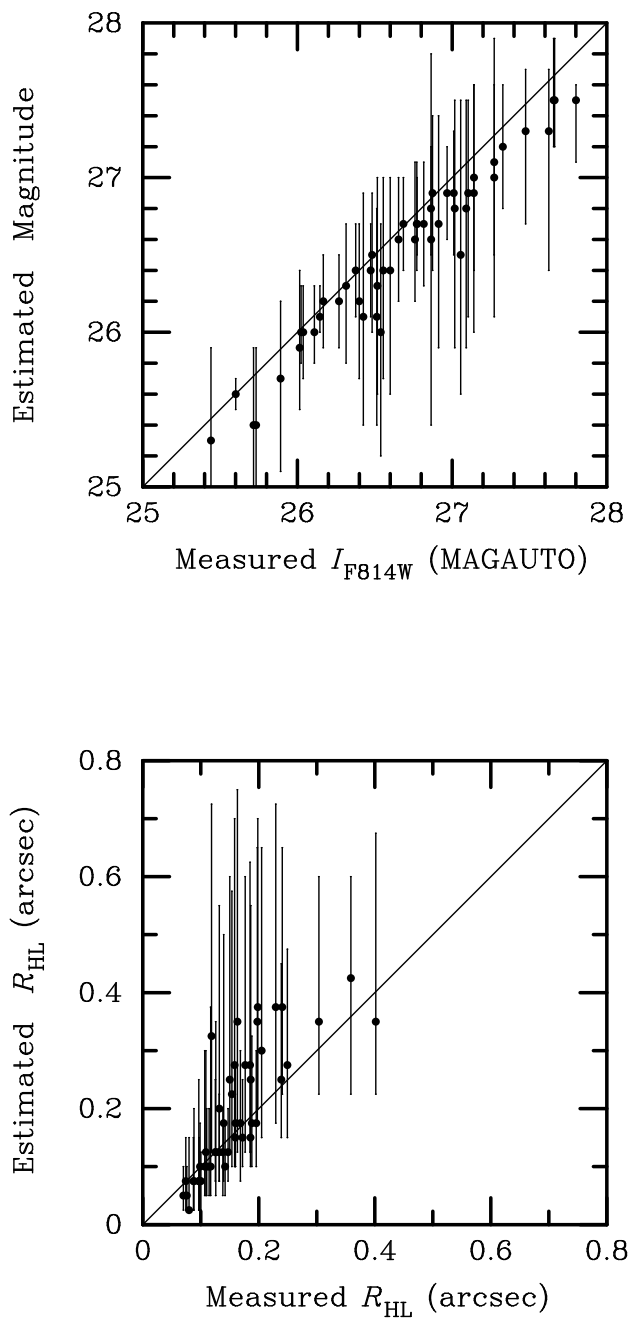}
\caption{Results of the Monte-Carlo simulation:
the relation between estimated total magnitudes by the simulation and measured
magnitudes  of our LAE sample (upper panel) and the relation 
between estimated half light radii by the simulation and measured
half light radii (upper panel). }
\end{figure}
\clearpage
\epsscale{1.0}


\begin{figure}
\plotone{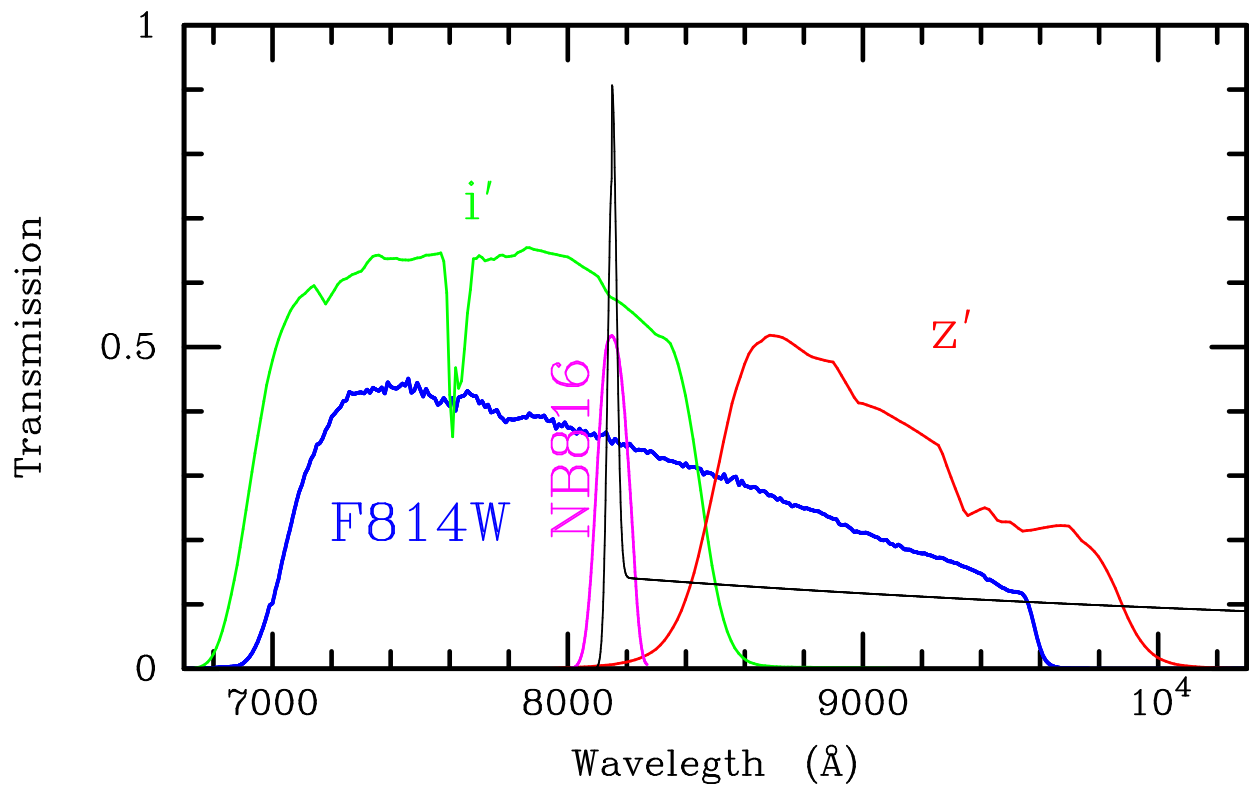}
\caption{Transmission curves for the filters used in our analysis; F814W (blue) for ACS/HST
and $i'$ (green) , $NB816$ (magenta), and $z'$ (red) for Suprime-Cam/Subaru. The CCD sensitivity is 
taken into account for each filter transmission curve.
Model spectrum (black) of a LAE at $z\approx 5.7$ with a rest-frame EW of 30\AA{} is also plotted.}
\end{figure}
\clearpage


\begin{figure}
\plotone{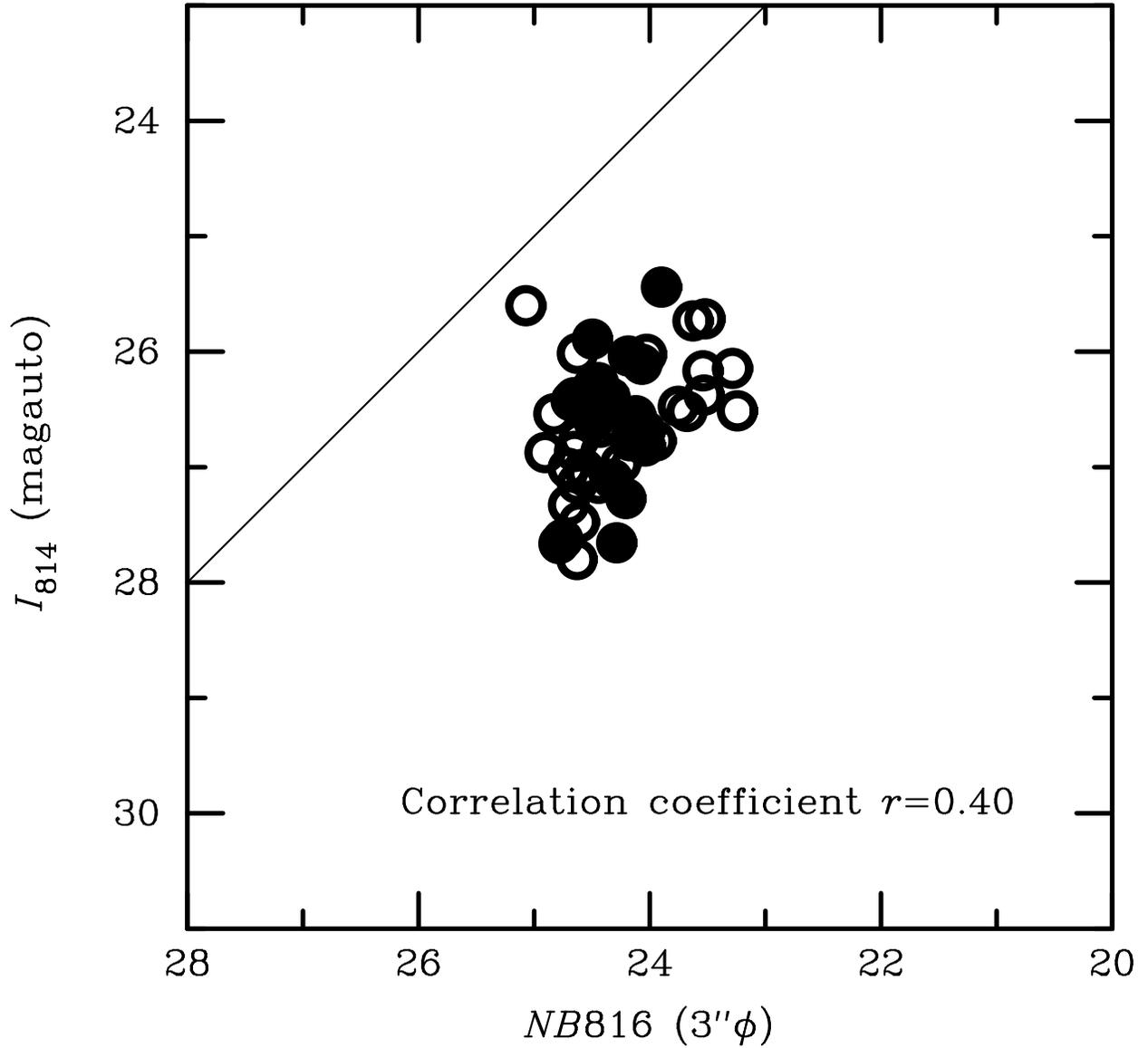}
\caption{Diagram between $I_{814}$ and $NB816$.
LAEs confirmed by our follow-up spectroscopy are shown by filled circles.
For the double-component LAEs, $I_{814}$ is the total magnitude of the two components.
Their $NB816$ magnitudes are measured with a 3-arcsec aperture that included the two components.
}
\end{figure}
\clearpage


\begin{figure}
\plottwo{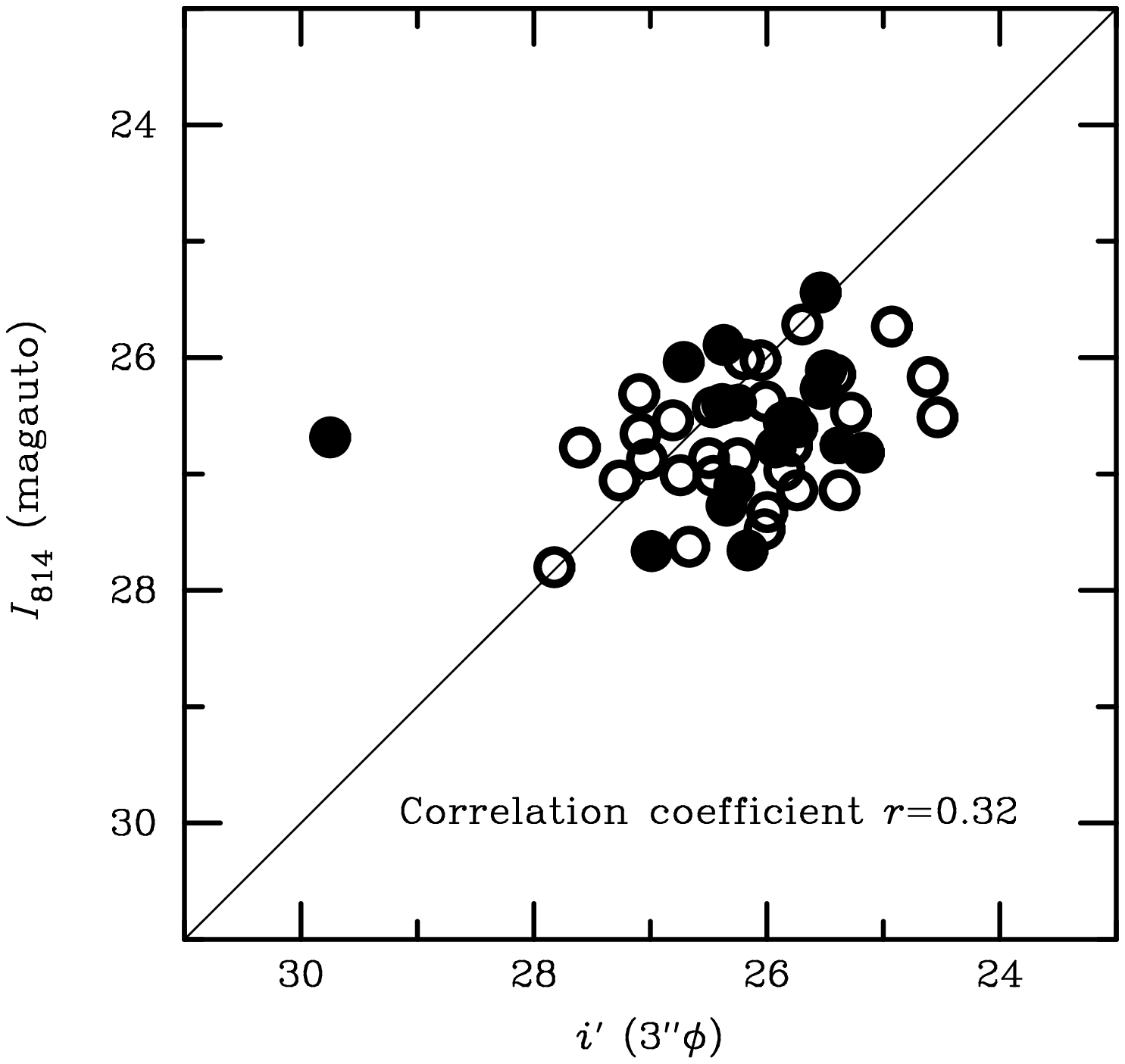}{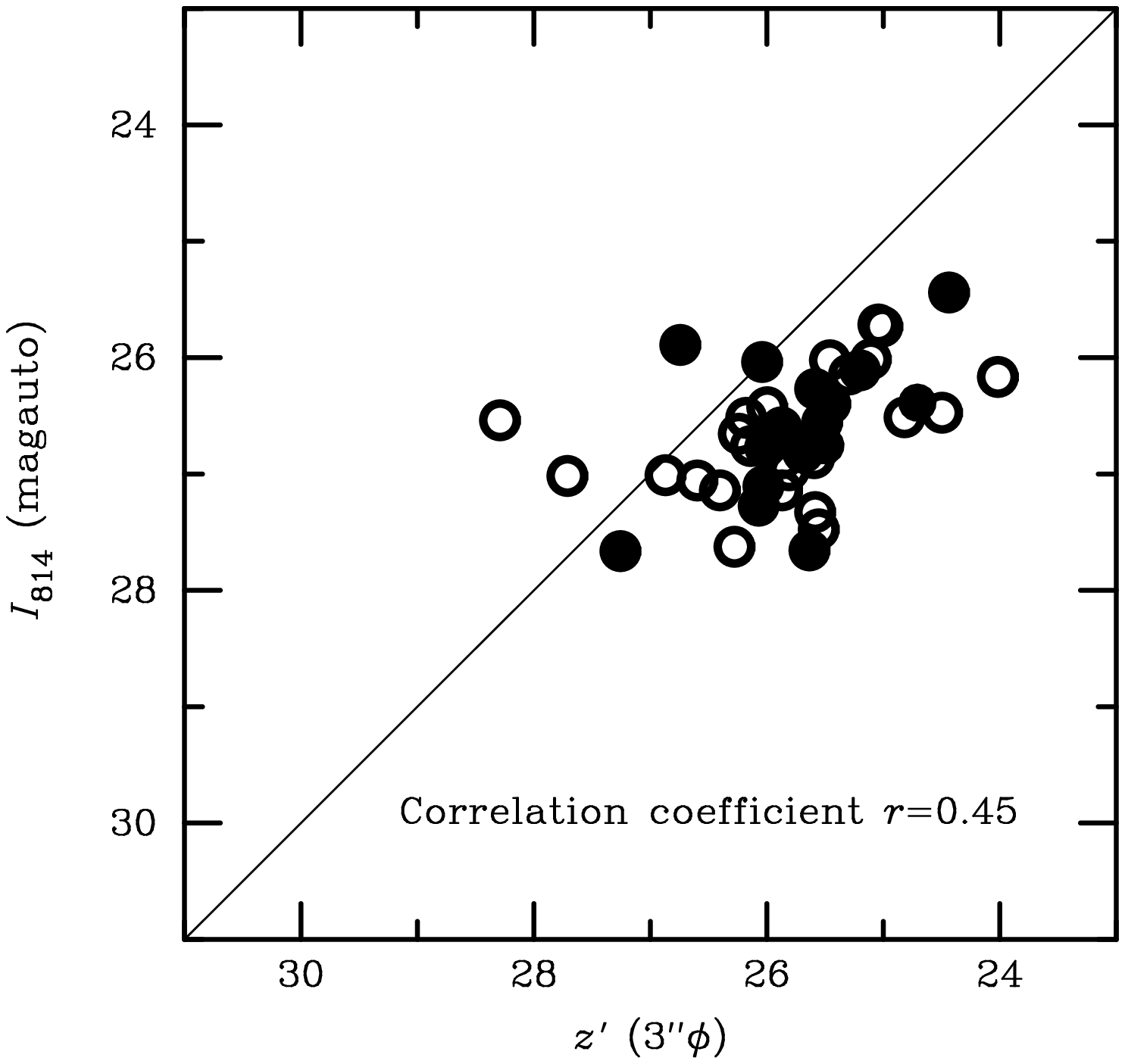}
\caption{Diagrams between $I_{814}$ and $i'$ (left), $z'$ (right).
LAEs confirmed by our follow-up spectroscopy are shown by filled circles.
For the double-component LAEs, $I_{814}$ is the total magnitude of the two components.
Their $i'$ and $z'$ magnitudes are measured with a 3-arcsec aperture that included the two components.
}
\end{figure}
\clearpage


\begin{figure}
\plotone{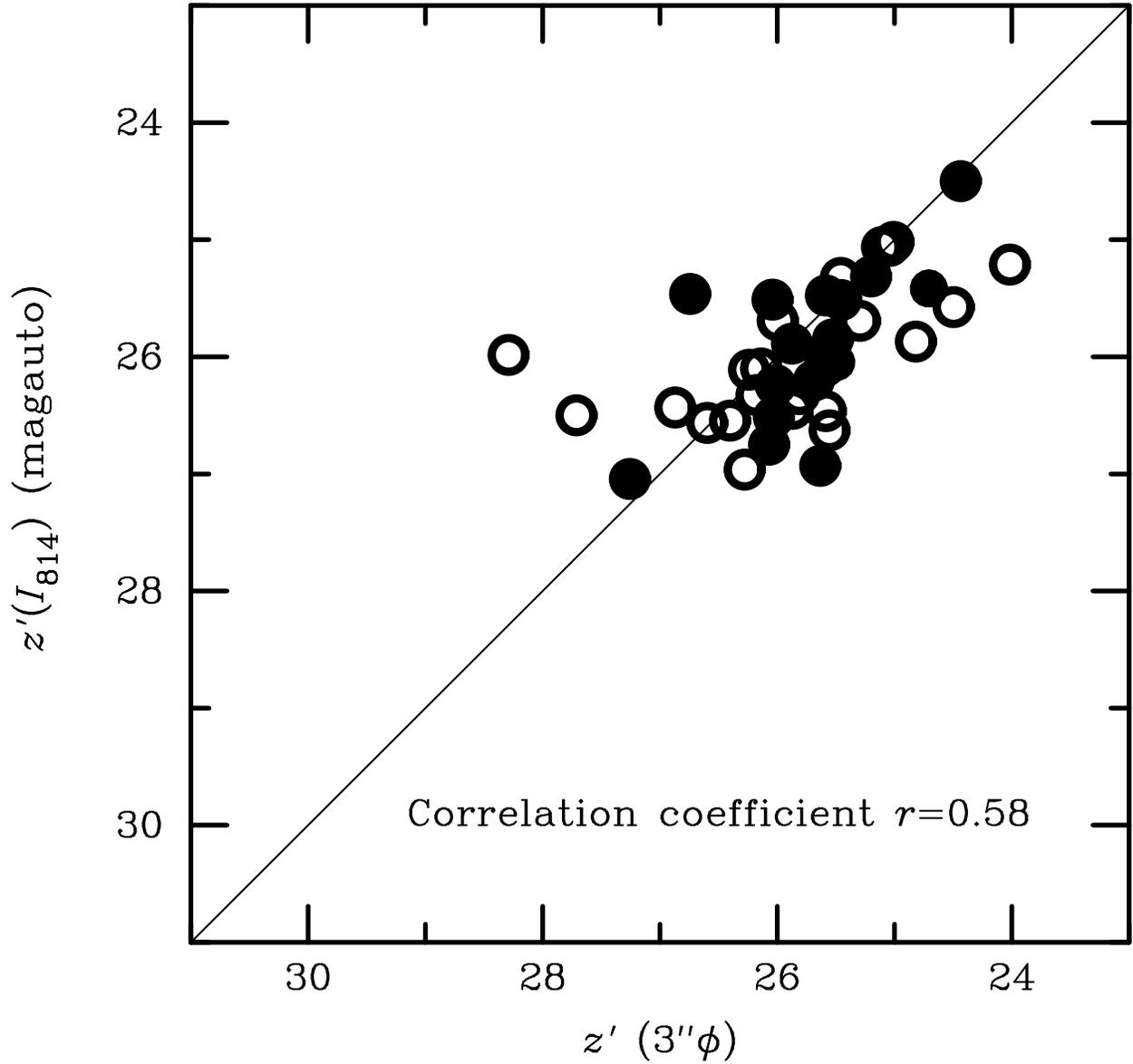}
\caption{Diagrams between the corrected $z$($I_{814}$) and $z'$.
LAEs confirmed by our follow-up spectroscopy are shown by filled circles.
For $z$($I_{814}$)  of the double-component LAEs, their total magnitudes of the two components are used.
Their Subaru $z'$ magnitudes are measured with a 3-arcsec aperture that included bot components.
}
\end{figure}
\clearpage

\begin{figure}
\plotone{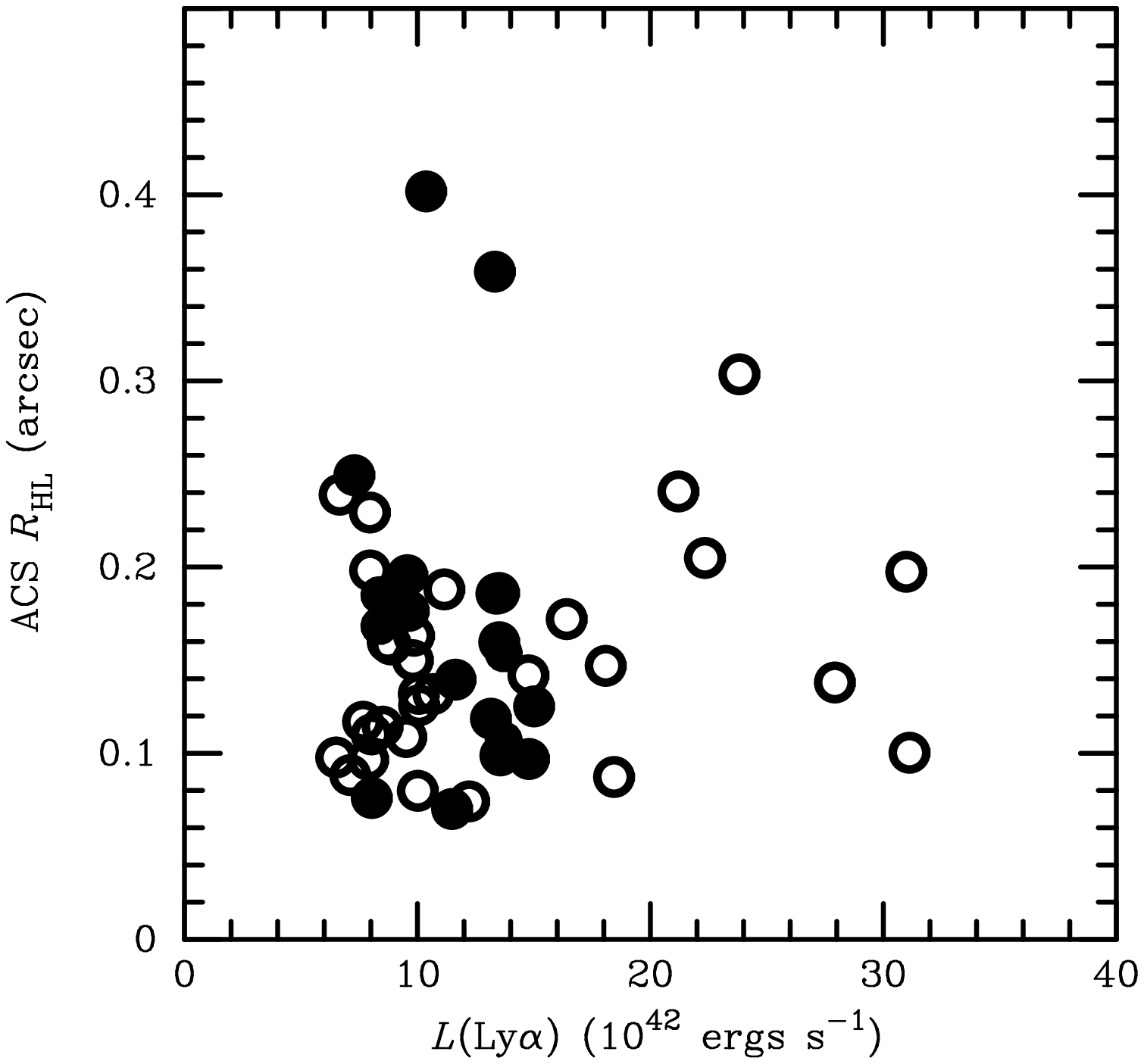}
\caption{Diagrams between $R_{\rm HL}$ and $L$(Ly$\alpha$).
LAEs confirmed by our follow-up spectroscopy are shown by filled circles.
For double-component LAEs $R_{\rm HL}$ is plotted for each component
while the same $L$(Ly$\alpha$) is adopted for the two components,
because they are not resolved in the ground-based images.
}

\end{figure}
\clearpage

\begin{figure}
\plotone{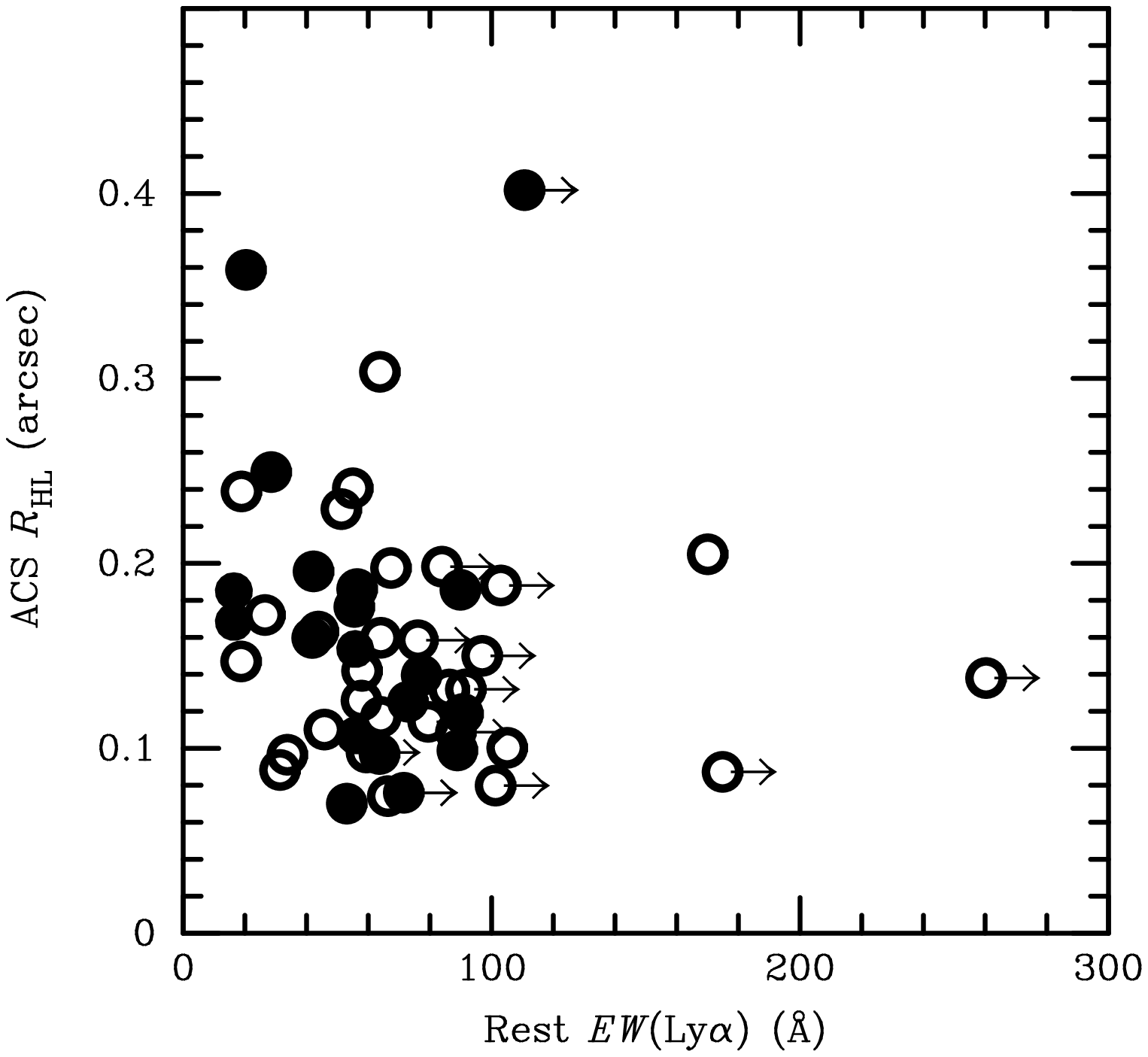}
\caption{Diagrams between $R_{\rm HL}$ and $EW$(Ly$\alpha$).
LAEs confirmed by our follow-up spectroscopy are shown by filled circles.
LAEs confirmed by our follow-up spectroscopy are shown by filled circles.
For double-component LAEs $R_{\rm HL}$ is plotted for each component
while the same $EW$(Ly$\alpha$) is adopted for the two components,
because they are not resolved in the ground-based images.
}
\end{figure}
\clearpage

\begin{figure}
\epsscale{0.7}
\plotone{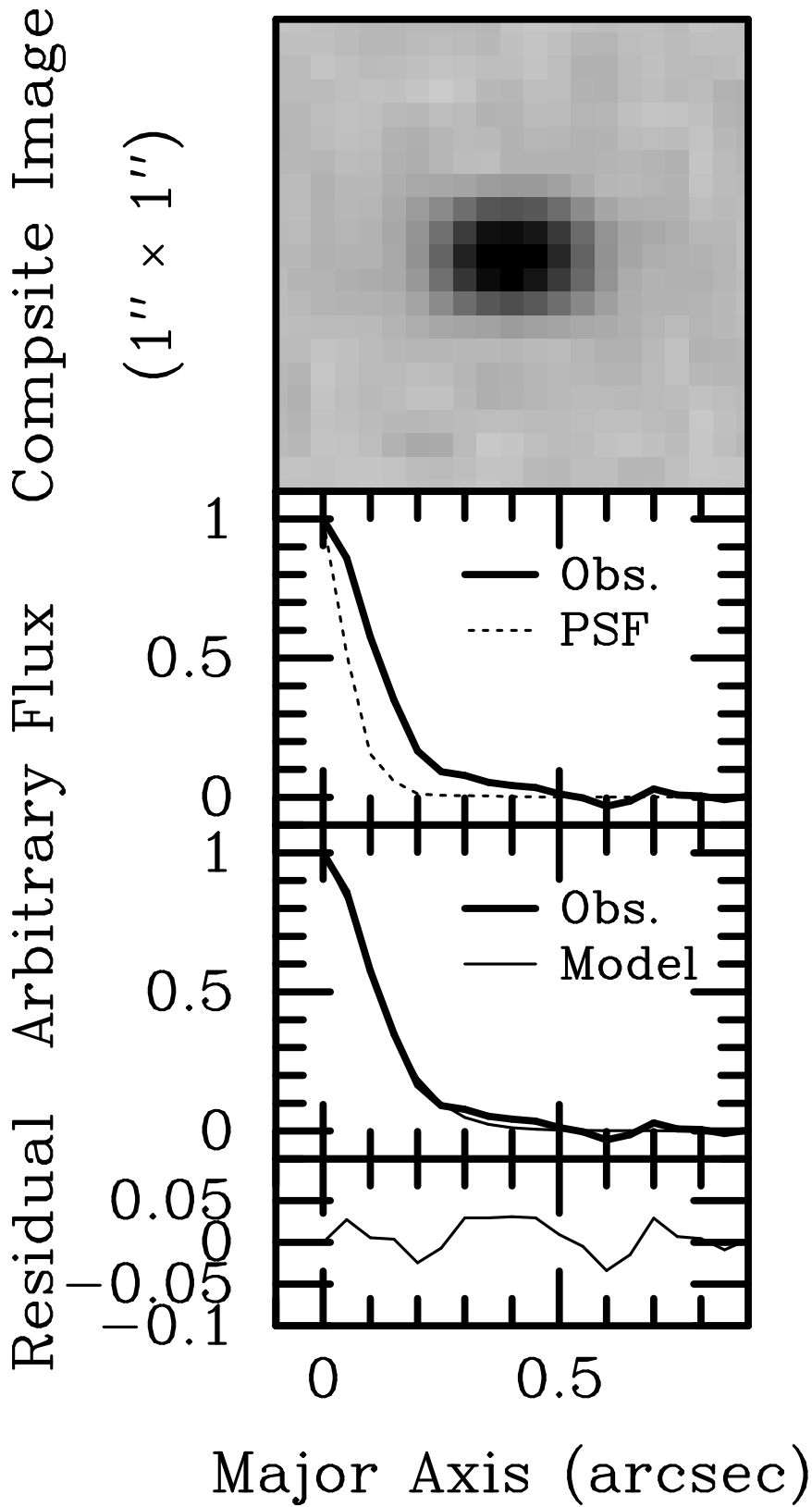}
\caption{Composite (stacked) ACS image of  all LAEs.
The surface brightness profiles (thick lines) , the model profiles (red lines) and residuals (thin likes)
are plotted.
The light profile of the PSF (dashed line) derived from stars in the ACS image is also indicated  for comparison.}
\end{figure}
\clearpage
\epsscale{1.0}

\begin{figure}
\plotone{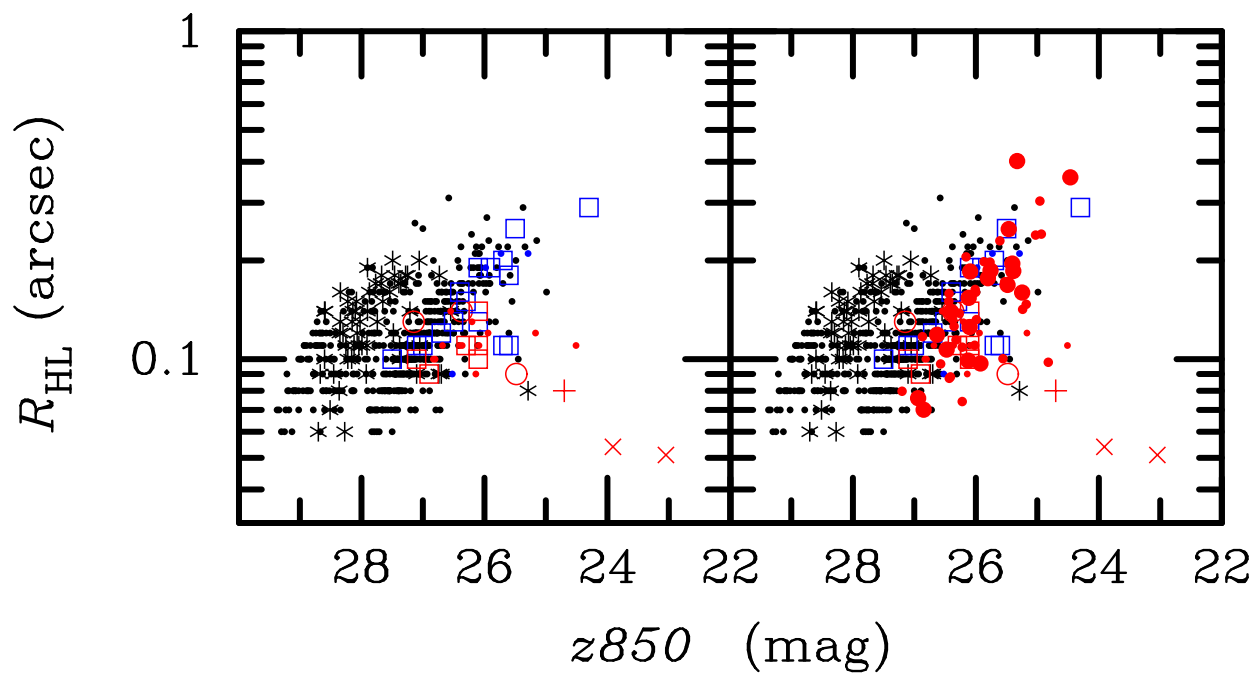}
\caption{Diagrams between $R_{\rm HL}$ and $z_{850}$ for $z\sim 6$.
The left panel is the same as the right panel except that our data is excluded for clarity.
Spectroscopically confirmed LAEs are shown by the red symbols and spectroscopically confirmed LBGs without
Ly$\alpha$ emission are shown by the blue symbols.
Our LAEs are shown by large red filled circles.
For the  double-component LAEs, each component is plotted.
See  Table 5 for explanation of the remaining symbols.
}
\end{figure}
\clearpage

\end{document}